\documentclass[utf8]{frontiersHLTHarxiv} 

\setcitestyle{square} 
\usepackage{url,hyperref,lineno,microtype,subcaption}
\usepackage[onehalfspacing]{setspace}
\usepackage{multirow}
\usepackage{tikz}
\usetikzlibrary{arrows,positioning,shapes.geometric}
\usepackage{algorithm}
\usepackage{algorithmic}
\usepackage{wrapfig}

\newcolumntype{?}{!{\vrule width 2pt}}

\usepackage{tablefootnote}
\usepackage{scrextend}

\newcommand{\ud}{\,\mathrm{d}}
\newcommand{\RR}{\mathbb{R}}

\newcommand{\NN}{\mathbb{N}}
\newcommand{\CC}{\mathbb{C}}

\newcommand{\ours}{\text{deppG}}

\def\keyFont{\fontsize{8}{11}\helveticabold }
\def\firstAuthorLast{Cicone, {et~al.}} 
\def\Authors{Cicone, Antonio\,$^{1}$ and Wu, Hau-Tieng\,$^{2,3,*}$}
\def\Address{$^{1}$Marie Curie fellow of the INdAM, DISIM, Universit\'a degli Studi dell'Aquila, via Vetoio 1, 67100, L'Aquila, Italy \\
$^{2}$Department of Mathematics and Department of Statistical Science, Duke University, Durham, NC, USA\\
$^{3}$Mathematics Division, National Center for Theoretical Sciences, Taipei, Taiwan}

\def\corrAuthor{Wu, Hau-Tieng}

\def\corrEmail{hauwu@math.duke.edu}

\begin{document}

\onecolumn
\firstpage{1}

\title[IHR and IRR from PPG via de-shape SST]{How {nonlinear-type time-frequency analysis} can help in sensing instantaneous {heart rate and instantaneous respiratory rate} from photoplethysmography in a reliable way}

\author[\firstAuthorLast ]{\Authors} 
\address{} 
\correspondance{} 

\extraAuth{}

\maketitle

\begin{abstract}
Despite the population of the noninvasive, economic, comfortable, and easy-to-install photoplethysmography (PPG), it is still lacking a mathematically rigorous and stable algorithm which is able to simultaneously extract from a single-channel PPG signal the instantaneous heart rate (IHR) and the instantaneous respiratory rate (IRR).
In this paper, a novel {algorithm called $\ours$} is provided to tackle this challenge. {$\ours$ is composed of two theoretically solid nonlinear-type time-frequency analyses techniques, the de-shape short time Fourier transform and the synchrosqueezing transform, which allows us to extract the instantaneous physiological information from the PPG signal in a reliable way.} 
To test its performance, {in addition to validating the algorithm by a simulated signal and discussing the meaning of ``instantaneous'',} the algorithm is applied to two publicly available batch databases, the Capnobase and the ICASSP 2015 signal processing cup. The former contains PPG signals relative to spontaneous or controlled breathing in static patients, and the latter is made up of PPG signals collected from subjects doing intense physical activities.
The accuracies of the estimated IHR and IRR are compared with the ones obtained by other methods, and represent the state-of-the-art in this field of research.
The results suggest the potential of {$\ours$} to extract instantaneous physiological information from a signal acquired from widely available wearable devices, even when a subject carries out intense physical activities.
\tiny
\keyFont{\section{Keywords:} de-shape short time Fourier transform, de-shape synchrosqueezing transform, photoplethysmography, instantaneous respiratory rate, instantaneous heart rate}
\end{abstract}

%

\section{Introduction}

Since being introduced in 1937 and its popularization in 1975 \cite{Aoyagi:1975,Nilsson2013}, photoplethysmography (PPG) has become an essential optical technique in healthcare and its mechanism has been extensively studied \cite{Mannheimer2007,Nilsson2013}. It is noninvasive, economical, comfortable and easy-to-install. In recent years, due to the advances in sensor technologies, different types of PPG signals have also become available via non-contact sensors \cite{Davlia_Lewis_Porges:2016,McDuff2016}. Furthermore, the PPG has become a standard equipped sensor in diverse mobile devices for healthcare, an important component in the internet of things \cite{Swan2012}, and in non-standard applications like monitoring the hemodynamics under the hyper- or microgravity environment \cite{Blanik2007}, in music therapy \cite{Shin2014} and many others.

The PPG contains extensive physiological dynamics information, such as peripheral oxygen saturation, autonomic nervous system status, cardiac and respiratory dynamics and hypovolemic status \cite{Nilsson2013}.
In recent decades, based on PPG plain statistics, several indices have been proposed and extensively applied for clinical needs. Examples include but are not limited to \cite{Nilsson2013} the heart rate and respiratory rate monitoring, pleth variability index for the fluid status assessment, surgical pleth index for stress evaluation and sleep apnea detection.

In more recent years, different scientific communities have tackled the problem of also learning hidden physiological information from a PPG. In particular, researchers have focused their attention on fine physiological dynamics like heart rate variability (HRV) and respiratory rate variability (RRV) that traditionally were studied directly from electrocardiograms or breathing flow signals.
The reason behind this effort is that reliably and accurately extracting fine physiological information from a PPG signal would open the door to the development and popularization of a new generation of medical care based on leisure time equipment.

However, analyzing the HRV and RRV from the PPG is not an easy task, particularly when we only have one channel PPG sensor. The main difficulty in achieving this research goal is the need to extract the heart rate and the respiratory rate from the PPG {\em in the instantaneous sense}, instead of simply getting average values over a time window. The identification of the instantaneous heart rate (IHR) and the instantaneous respiratory rate (IRR) from a PPG recording is challenging for standard signal processing tools due to the following two factors.
The first challenge comes from the time-variability in the heart and respiratory rate, and the second challenge is a result of the non-sinusoidal PPG oscillation. The time-variability in the heart and respiratory rate broadens the spectrum, and the non-sinusoidal oscillation inevitably mixes up the spectral information for the cardiac and respiratory activity \cite{lin2016waveshape}. The broadened and mixed up spectrum prohibits us from applying the standard signal processing techniques. The problem is even more challenging, since the signal is often contaminated by nonstationary uncertainties, like noise and motion artifacts, particularly when PPG signals are recorded during every day activities.

In the past few years, several methods have been proposed to address this challenge.
For the IHR, methods include time-frequency (TF) analyses \cite{Gil2010,Wu_Lewis_Davila_Daubechies_Porges:2015,Mullan2015}, adaptive filtering \cite{Yousefi2014,Boloursaz2015,Khan2015,murthy2015,Schack2015}, Kalman filter \cite{Frigo2015}, sparse spectrum reconstruction \cite{Zhang2015}, blind source separation \cite{Wedekind2015}, a Bayesian approach \cite{Sun2015,SandeepJarChakraborti2015}, correntropy spectral density (CSD) \cite{Garde2014}, empirical mode decomposition (EMD) \cite{ZhangLiuZhang2015a}, model fitting \cite{Wadehn2015}, deep learning \cite{Jindal2016}, fusion approaches \cite{Temko2015,Zhu_Tan:2015}, etc.
For the IRR, efforts include TF analysis \cite{Chon2009,Orini_Pelaez-Coca:2011,Dehkordi2015}, sparse signal reconstruction \cite{Zong2015,Zhang2016}, neural network \cite{Johansson2003}, modified multi-scale principal component analysis \cite{Madhav2013}, independent component analysis \cite{Zhou2006}, time-varying autoregressive regression \cite{LeeChon:2010a,LeeChon:2010b}, fusion approaches \cite{Karlen2013,Cernat2015}, pulse-width variability \cite{Lazaro2013,Ungureanu2014}, CSD \cite{Pelaez-Coca2013,Garde2014}, EMD \cite{garde2013empirical}, a Bayesian approach \cite{Pimentel2015,Zhu_Pimentel:2015}, etc.
While the above algorithms focus on either IHR or IRR, only a few ad-hoc algorithms are considered to extract simultaneously the IHR and IRR, like \cite{garde2013empirical,Garde2014}.
Except for a few algorithms, most of the above methods are based on multiple-channel signals. For example, the method based on the pulse wave transit time \cite{Johansson2006} needs at least one PPG channel and one ECG channel. At least two PPG channels are needed in the adaptive filtering and blind source separation method.

\begin{figure}[ht]
\centering
	\includegraphics[width=.9\linewidth]{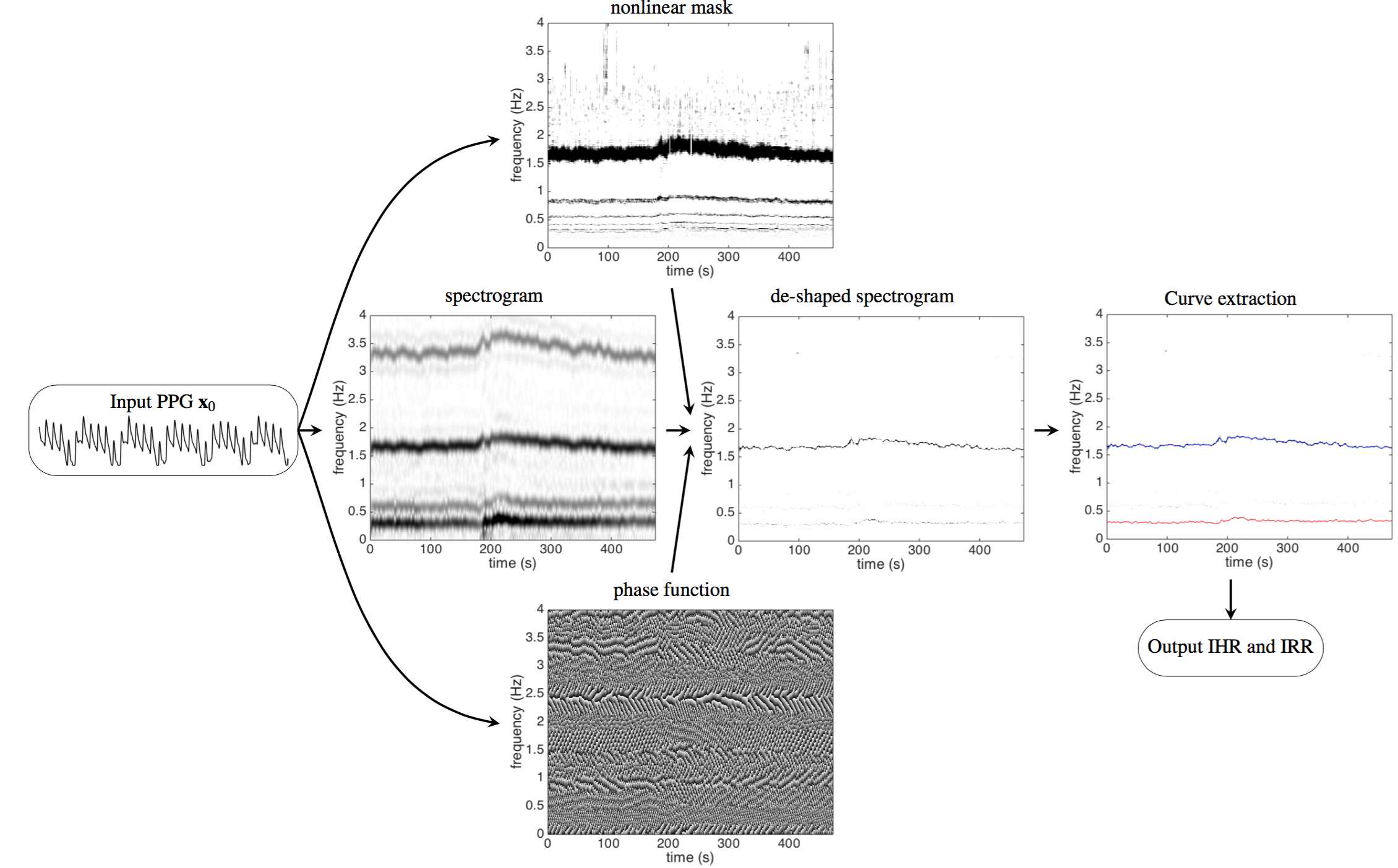}
    \caption{\label{FlowChart}The flow chart of the proposed algorithm, $\ours$, to extract the instantaneous heart rate (IHR) and instantaneous respiratory rate (IRR) from the recorded PPG signal. A typical recorded PPG signal lasting for 20 seconds is shown on the left. The short time Fourier transform (STFT), and hence the spectrogram, of the input PPG signal are then evaluated.
The intensity of the spectrogram at a point $(t,\xi)$ in the time-frequency plane indicates how strong the signal oscillates at time $t$ and frequency $\xi$. The dark curve around 1.6Hz represents the IHR, while the gray curve around 3.2Hz (and 4.8Hz, 6.4Hz, etc. The frequency axis above 4Hz is not shown) comes from the non-sinusoidal oscillation of the cardiac activity. Similarly, the dark curve around 0.3Hz represents the IRR, while the gray curve around 0.6Hz comes from the non-sinusoidal oscillation of the respiratory activity.
With the STFT and the spectrogram of the PPG signal, the nonlinear mask is then designed from the spectrogram and the phase function is determined from the STFT. The intensity of the phase function at a point $(t,\xi)$ in the time-frequency plane indicates the angle of the complex value of the STFT at time $t$ and frequency $\xi$, which ranges from $0$ to $2\pi$.
By applying the nonlinear mask and the phase function of the STFT to the spectrogram, the spectrogram is improved and we obtain the de-shaped spectrogram. The darker curve around 1.6 Hz represents the IHR and the lighter curve around 0.3 Hz represents the IRR. The curves corresponding to the IHR and IRR are extracted from the de-shaped spectrogram, which are shown as the red and blue curves respectively superimposed on the de-shaped spectrogram.}
\end{figure}

In this work, we propose a new signal processing technique with deep mathematical roots: the {\em $\ours$ algorithm}, which is based on a \textit{nonlinear masking technique}. This method allows the extraction of IHR and IRR simultaneously from a single channel PPG signal, even in presence of high level of noise and motion artifacts.
We {validate it on a semi-real simulated PPG database, and discuss the meaning of ``instantaneous''. Then apply $\ours$ to} two publicly available databases, CapnoBase benchmark database (\url{http://www.capnobase.org}) and ICASSP 2015 signal processing cup (\url{http://www.zhilinzhang.com/spcup2015/}), and compare the results with the ones obtained by other methods.
The proposed $\ours$ algorithm is composed of two steps. In the first one, given a recorded PPG signal, we compute its spectrogram using the short time Fourier transform (STFT). In order to accurately and robustly extract the IHR and IRR information from the PPG signal, we sharpen the spectrogram both by applying a novel data adaptive nonlinear mask \cite{lin2016waveshape}, and by taking into account the phase information of the STFT of the recorded PPG \cite{Daubechies_Lu_Wu:2011,Chen_Cheng_Wu:2014}. We call the resulting information the \textit{de-shaped spectrogram}.
Thanks to this mask, the {instantaneous physiological} information contained in the PPG is preserved and enhanced in the de-shape spectrogram at the same time. We point out that the novel nonlinear mask is data adaptive in the sense that it is determined directly from the recorded PPG.
Second, we apply a curve extraction technique to derive the IHR and IRR curves from the de-shaped spectrogram.
The $\ours$ algorithm structure is illustrated in Figure \ref{FlowChart}.

The rest of the paper is organized as follows.
A detailed description of the proposed $\ours$ algorithm, {including the numerical implementation for the reproducibility purpose,} is given in Section \ref{Section:Methods}, together with a new non-harmonic model which describes a generic PPG signal. In Section \ref{Section:Results}, {$\ours$ is validated on a simulated PPG database, a discussion of the meaning of ``instantaneous'' is provided, and} we apply $\ours$ to the CapnoBase benchmark database and ICASSP 2015 signal processing cup datasets, and compare results with the ones obtained by other techniques proposed so far in the literature. We conclude this paper with a discussion section.
In the online Supplementary Information (SI), 
we report extra information regarding the results obtained applying this technique to the CapnoBase and  ICASSP database.

\section{Methods}\label{Section:Methods}

\subsection{Mathematical Model}
We believe that to properly analyze a PPG signal, it is fundamental to first develop a suitable model able to replicate the main features of such a signal.
With this aim, we start by making an important observation regarding the physiological dynamics we want to model. The cardiac dynamics in a PPG signal could be well understood by taking the Lambert-Beer law into account \cite{Mannheimer2007}. On the other hand, there are several sources of the respiratory dynamics in the PPG, for example, the respiratory-induced intensity variations (RIIV) and the pulse-wave transit time \cite{Mannheimer2007,Nilsson2013}. These dynamics are oscillatory in nature and inherited from the oscillatory central drivers of the cardiac and respiratory activities. Since writing down a thoughtful differential equation for the PPG might not be feasible due to such complicated physiological dynamics, we choose to model the PPG by a statistics-based phenomenological model that mainly describes the empirical observations of the PPG.

As the first idea for a model, we would be tempted to use a \textit{harmonic model}. However, the PPG cannot be properly modeled by a harmonic model, like sine or cosine functions, due to at least three limitations.
First, in a harmonic model, the frequency, or the period of each oscillation, is assumed to be fixed. This limits the ability of the model to capture the irregular oscillation of the PPG. Instead, the notion of \textit{time-varying frequency}, or \textit{instantaneous frequency}, is needed. As the name suggests, the time-varying frequency captures the fact that the PPG oscillatory periods vary over time. Physiologically, the time-varying frequency of the cardiac dynamics captures the IHR, and the time-varying frequency of the respiratory dynamics captures the IRR.
Second, the amplitude is assumed to be fixed in a harmonic model. Note that the amplitude of each PPG oscillation is directly related to the blood pressure, which is not constant over time. Therefore, a \textit{time-varying amplitude} is needed to better capture the PPG behavior.
Third, the oscillation is assumed to be sinusoidal in the harmonic model. It is well known that the systolic and diastolic phases of the cardiac activity do not last the same amount of time \cite{Guyton:2000}. A similar fact holds for the respiratory activity \cite{Guyton:2000}. Therefore, we need to consider a \textit{non-sinusoidal shape} to model the oscillations in the PPG.
In summary, the harmonic model needs to be substituted by one that is able to capture at least three characteristic aspects of a PPG signal  -- time-varying frequency, a time-varying amplitude and a non-sinusoidal shape.

A well-studied substitute for the harmonic model that can work well in this context is the \textit{adaptive non-harmonic model} (ANHM) \cite{Wu:2013,lin2016waveshape}. A signal satisfies the ANHM, if it is oscillatory with a time-varying frequency, a time-varying amplitude and non-sinusoidal oscillations. Mathematically, it is expressed as
\begin{equation}
f(t)=\sum_{k=1}^Ka_k(t)s_k(\phi_k(t)),
\end{equation}
where $a_k(t)s_k(\phi_k(t))$ is the $k$-th oscillatory component in the recorded signal $f(t)$, called the \textit{intrinsic-mode-type function}, $a_k(t)$ is the time-varying amplitude, the $\phi_k(t)$ is a monotonically increasing function whose derivative is the time-varying frequency and $s_k$ is a $1$-periodic function that captures the non-sinusoidal shape. We call the derivative of $\phi_k(t)$ and $s_k$ the \textit{fundamental frequency} and the \textit{wave-shape function} of the $k$-th intrinsic-mode-type function respectively. We represent the single-channel PPG signal by the ANHM with $K=2$, where the first intrinsic-mode-type function, $a_1(t)s_1(\phi_1(t))$, models the respiratory dynamics and the second intrinsic-mode-type function, $a_2(t)s_2(\phi_2(t))$, models the cardiac dynamics.
This model is still too general. We need to incorporate in it some physiological knowledge regarding the phenomena we want to describe. The first observation regards the variation of IHR and IRR. Even though they are not constant, they do not change suddenly based on the physiological homeostasis assumption, unless there is any unexpected impact from the outside system. We thus assume that the IHR and IRR vary slowly over time. This implies that both the time-varying frequencies and the time-varying amplitudes vary slowly over time in this model. Under normal physiological circumstances, we further assume that the fundamental frequency of the cardiac dynamics in the ANHM is larger than the one of the respiratory dynamics. Under all these assumptions, the ANHM can be properly analyzed from a mathematical standpoint.
We refer interested readers to \cite{Wu:2013,lin2016waveshape} for more technical details on this model, like the time-varying wave-shape function, and its mathematical analysis.

\subsection{Proposed algorithm: $\ours$}
Analyzing the ANHM, and hence the PPG, is generally a challenging problem from a signal processing viewpoint. Due to the time-varying nature of the amplitudes and fundamental frequencies, if we try using standard techniques to study PPG signals from a TF point of view, we discover that the spectrum of each component of a PPG signal is broad. The problem is further complicated by the non-sinusoidal nature of the wave-shape function which makes each spectrum become even broader. To complete the picture, we have to consider that having a broad spectrum for each oscillatory component implies they will interact with each other in the frequency domain. All of these facts together explain why analyzing a PPG signal is almost an impossible problem for standard signal processing methods. Figure \ref{fig:MotionPPG_TFR} shows an illustration of this challenge.
To be more specific, there are at least two signal processing challenges we need to resolve, if we want to simultaneously extract the IHR and IRR from a single-channel PPG signal. The first challenge is to deal with the broad spectrum determined by the time-varying nature of the IHR and IRR{, and hence obtain the instantaneous information}. The second challenge is to take care of the interaction between the cardiac and respiratory activities that is caused by non-sinusoidal wave-shape functions.

{ 
Below, we detail the proposed $\ours$ algorithm that handles the above mentioned challenges. A summary of the algorithm is shown in Figure \ref{FlowChart}. For readers with interest in the theoretical analysis and related mathematical theorems, we refer them to \cite{lin2016waveshape}.}

\begin{figure}[ht]
\centering
\includegraphics[width=.9\linewidth]{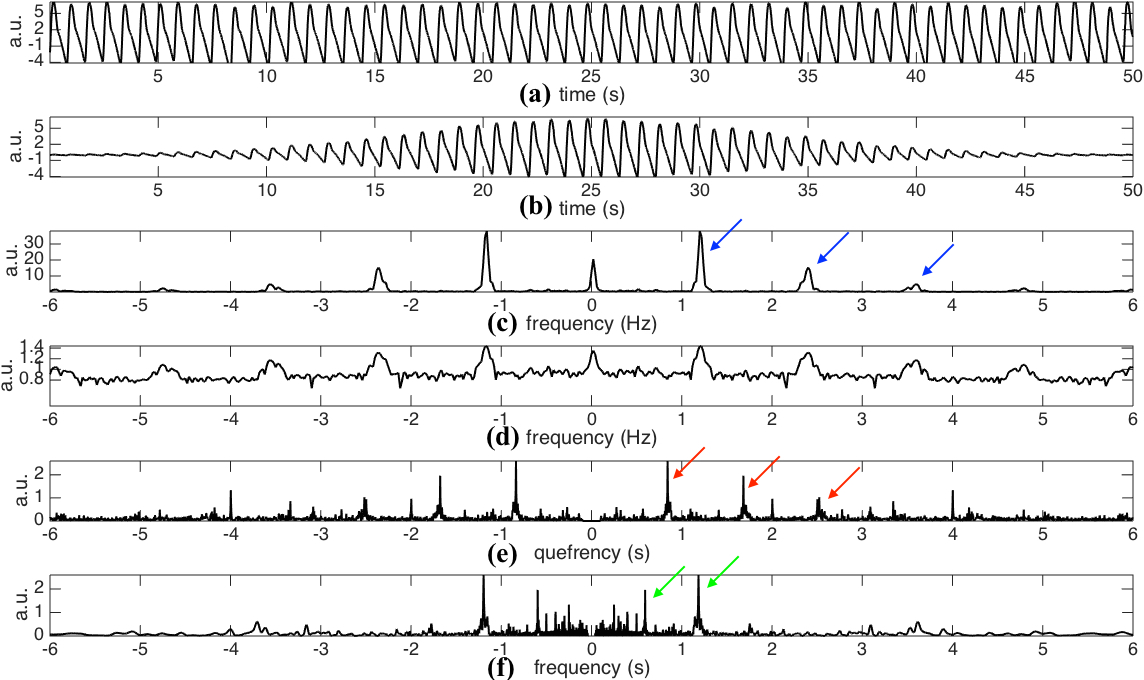}
\caption{(a) The photoplethysmogram (PPG) signal of the dataset \texttt{0031\textunderscore 8min} in the Capnobase database lasting for 50 seconds. The signal is shifted by $1$ to simulate the trend commonly encountered in real data. (b) The ``windowed'' PPG signal that is generated by multiplying the PPG signal by a Gaussian window centered at the 25-th second. (c) The power spectrum of the windowed PPG signal shown in (b). The fundamental frequency, $1.2$Hz, and its multiples, $1.2\times 2, 1.2\times3,\ldots$, etc, are indicated by the blue arrows. Note that the magnitude varies from peak to peak, which depends on the non-sinusoidal oscillation of the PPG signal. Also note that this power spectrum is the spectrogram at the 25-th second. (d) The $0.1$ power of the power spectrum shown in (c). It is clear that the magnitudes of all peaks become more uniform after taking the fractional power. We could thus view it as a periodic function with the ``period'' exactly the same as the fundamental frequency of the PPG signal. (e) The spectrum of the (d) that contains the {\em fundamental period} information, which is the inverse of the fundamental frequency, of the PPG signal at the 25-th second. Specifically, that the peaks indicated by the red arrows located at $1/1.2, 2/1.2,\ldots$, etc, are associated with the fundamental period and its multiples. (f) The nonlinear mask determined by inverting the quefrency axis by sending a nonzero quefrency $q$ to $1/q$. We could clearly see that the peaks indicated by the red arrows in (e) become the peaks indicated by the green arrows (f). By a direct algebraic calculation, the peaks indicated by the green arrows correspond to the fundamental frequency, $1.2$ Hz, and its divisions, $1.2/2,1.2/3,\ldots$, etc.}
\label{Figure:MaskDesign}
\end{figure}

\subsubsection{Step1: Handle the first challenge by the short-time Fourier transform}
 

We find the ``local spectrum'' of the PPG by computing the Fourier transform of a small piece of the PPG cut out by a chosen sliding window function. By concatenating the local spectra together according to time, we obtain the time-varying spectrum that is commonly named the \textit{spectrogram}. 
{Mathematically, the STFT of $f\in \mathcal{S}'$ with a chosen window function $h\in\mathcal{S}$, where $\mathcal{S}$ is the Schwartz space and $\mathcal{S}'$ is the space of tempered distributions, is defined as
\begin{equation}
V^{(h)}_f(t, \xi) = \int f(\tau) h(\tau-t)e^{-i2\pi \xi (\tau-t)} \ud \tau\,,\label{eq: stft1}
\end{equation}
where $t\in\RR$ indicates time and $\xi\in\RR$ indicates frequency. We call $|V^{(h)}_f(t, \xi)|^2$ the spectrogram of the signal $f$.}
The spectrogram thus captures the time-varying fundamental frequencies of each component contained in the PPG, including the IHR and IRR.

See Figure \ref{Figure:MaskDesign} for an example. For a given the PPG signal shown in Figure \ref{Figure:MaskDesign} (a), we can compute its spectrogram using the STFT. The spectrogram, as explained before, provides for each instant of time the ``local spectrum'' of the PPG. In particular, if we consider the instant corresponding to 25 seconds, the corresponding local spectrum is obtained by first multiplying the PPG signal by a Gaussian window centered at the 25-th second, whose outcome is shown in Figure \ref{Figure:MaskDesign} (b), and then applying the Fourier transform to such a signal. The outcome is the local spectrum corresponding to the 25-th second of the signal, which is shown in Figure \ref{Figure:MaskDesign} (c).

From this local spectrum, we discover that the non-sinusoidal shape function, shown in Figure \ref{Figure:MaskDesign} (a), has a periodic spectrum. Specifically, in addition to the fundamental frequency (the peak at about 1.2 Hz), in the local spectrum shown in Figure \ref{Figure:MaskDesign} (c), we have multiples of the fundamental frequency indicated by the blue arrows. Note that the multiples would have different strengths, which depend on the non-sinusoidal shape function.
By taking a small fractional power of the local spectrum, the strengths of the multiples become more ``uniform'', as plotted in Figure \ref{Figure:MaskDesign} (d). Therefore, we can focus on the periodic structure.

\subsubsection{Step 2: short-time cepstrum transform and the nonlinear mask}

{While the STFT provides a fruitful information about the cardiac and respiratory oscillation, however, it is not an easy task to extract the IHR and IRR information from the spectrogram computed using STFT due to the uncertainty principle of the STFT and the second challenge. The interference of the spectral information from cardiac and respiratory activities due to their non-sinusoidal shape determines the coexistence in the spectrogram of both IHR, IRR, and their multiples. This phenomenon can be clearly seen in the spectrograms shown in Figures \ref{FlowChart} and \ref{fig:MotionPPG_TFR}.

To handle the second challenge,} the main trick is to take the Fourier transform {of the small fractional power of the local spectrum (shown in \ref{Figure:MaskDesign} (d)). By doing so, we} get the information about the oscillatory period and its multiples in what is called the \textit{quefrency} domain in the literature \cite{oppenheim2004frequency}. See Figure \ref{Figure:MaskDesign} (e) for an example. Specifically, the peaks indicated by the red arrows represent the oscillatory period and its multiples. The terminologies ``quefrency'' is invented by interchanging the consonants of the first part of the word ``frequency'' in order to emphasize that the information is not in the original time domain but is related to the frequency domain \cite{oppenheim2004frequency}. By definition, the quefrency has the same unit as time; in our analysis example, the unit is second. In the signal processing field, the function defined on the quefrency domain is called the {\em cepstrum} \cite{oppenheim2004frequency}. We then repeat such analysis for each instant of time to produce the local cepstrum at every second. By concatenating the local cepstrum together according to time, we get the \textit{short-time cepstrum transform} (STCT) of the PPG \cite{lin2016waveshape}. 
{Mathematically, the STCT is evaluated by
\begin{equation}
C^{(h,\gamma)}_f(t, q) := \int |V^{(h)}_f(t, \xi)|^\gamma e^{-i2\pi q \xi} \ud \xi,
\label{eq: rceps1}
\end{equation}
where $\gamma>0$ is sufficiently small and $q\in\RR$ is the quefrency.}

Thanks to the STCT, we are ready to design the newly proposed nonlinear mask.
Using the fundamental relationship between frequencies and periods, periods are the inverse of the frequencies, the nonlinear mask is obtained by inverting the quefrency axis of the STCT. Note that by inverting the period and its multiples, this nonlinear mask contains the fundamental frequency and its divisions. See Figure \ref{Figure:MaskDesign} (f) for an illustration. 
{Mathematically, this  nonlinear mask for the spectrogram is given by
\begin{equation}
U_f^{(h,\gamma)}(t,\xi):=C_f^{(h,\gamma)}(t,\mathcal{I}\xi),\label{Definition:NonlinearMask}
\end{equation}
where $\xi>0$ is in the unit of Hz and $U_f^{(h,\gamma)}(t,\cdot)$ is in general a distribution defined on $\RR\times \RR^+$.}

\subsubsection{Step 3: Handle the second challenge by applying the nonlinear mask}
As a result, the common component between the nonlinear mask and the spectrogram is only the fundamental frequency.
Therefore, by multiplying the nonlinear mask with the spectrogram, we ``filter out'' the spectral information associated with the non-sinusoidal shape function, and preserve the fundamental frequency information.
It is clear that the nonlinear mask just defined depends mainly on the PPG itself, so it is \textit{data adaptive}. This adaptivity feature of the nonlinear mask design allows the proposed method to, at the same time, automatically preserve and enhance the {instantaneous} information contained in the PPG signal regarding the IHR and IRR.
{Mathematically, this nonlinear mask satisfies
\begin{equation}
\label{eq:W}
W^{(h,\gamma)}_f(t, \xi) := V^{(h)}_f(t,\xi)U^{(h,\gamma)}_f(t, \xi),
\end{equation}
where $\xi>0$ is interpreted as frequency and $W^{(h,\gamma)}_f$ is defined on $\RR\times \RR^+$.}

\subsubsection{Step 4: Refine the first challenge by sharpening the time-frequency representation}

It has been shown in \cite[Theorem 3.6]{lin2016waveshape} that if we model the PPG signal by the ANHM, the IHR and IRR information shows up as two curves in the nonlinearly masked spectrogram.
To enhance the sharpness of these two curves, we apply another nonlinear transform technique called the synchrosqueezing transform \cite{Daubechies_Lu_Wu:2011,Chen_Cheng_Wu:2014}.
The synchrosqueezing transform sharpens the curves by taking the phase information of the STFT into account. Specifically, each spectrogram coefficient is nonlinearly reallocated to a new location that is determined by the phase information of the STFT. It has been shown in \cite{Daubechies_Lu_Wu:2011,Chen_Cheng_Wu:2014} that when a signal, like the PPG, is modeled by the ANHM, the spectrogram coefficients will be reallocated and the IHR and IRR curves will be enhanced. {The robustness of the synchrosqueezing transform to different kinds of noise has been studied in \cite{Chen_Cheng_Wu:2014}, and it has been applied to the non-contact PPG signal analysis to obtain the IHR of the state-of-art quality \cite{Wu_Lewis_Davila_Daubechies_Porges:2016}.
Mathematically, the synchrosqueezing transform \cite{Daubechies_Lu_Wu:2011,Chen_Cheng_Wu:2014} is defined as as
\begin{equation}\label{definition:SSTW}
SW^{(h,\gamma,\upsilon)}_{f}(t,\xi)=\int_{\mathfrak{N}_\upsilon(t)} |W^{(h,\gamma)}_f(t,\eta)| \frac{1}{\alpha}g\left(\frac{|\xi-\Omega^{(h,\upsilon)}_f(t,\eta)|}{\alpha}\right)\ud \eta\,,
\end{equation}
where $\xi\geq0$, $g\in\mathcal{S}$ so that $g(\cdot/\alpha)/\alpha$ converges weakly to the Dirac measure supported at $0$ as $\alpha\to 0$, and $\Omega^{(h,\upsilon)}_f$ is the \textit{reassignment rule} determined by taking the phase information in the STFT into account via
\begin{equation}
\Omega^{(h,\upsilon)}_f(t,\xi):=
\left\{
\begin{array}{ll}
-\Im\frac{V_f^{(\mathcal{D}h)}(t,\xi)}{2\pi V_f^{(h)}(t,\xi)}&\mbox{ when }|V_f^{(h)}(t,\xi)|> \upsilon\\
-\infty&\mbox{ when }|V_f^{(h)}(t,\xi)|\leq \upsilon
\end{array}
\right.
\,,\label{RM:omega}
\end{equation}
$\mathcal{D}h(t)$ is the derivative of the chosen window function $h\in\mathcal{S}$, $\Im$ means the imaginary part, and $\upsilon>0$. The threshold $\upsilon$ helps avoid instability in the computations when $|V^{(h)}_f(t,\xi)|$ is small. 
}
We call the outcome of the nonlinearly masked and sharpened spectrogram the \emph{de-shaped spectrogram}.
The readers can see examples of such de-shaped spectrograms in Figure \ref{FlowChart} and Figure \ref{fig:MotionPPG_TFR}.

\subsubsection{Step 5: Curve extraction to estimate the IHR and IRR}

Finally, since in the PPG de-shaped spectrogram the IHR and IRR show up as sharp curves, we could apply any curve extraction technique to estimate the IHR and IRR.
For simplicity, in this work, we use an intuitive optimization-based curve extraction algorithm to extract such curves \cite{Chen_Cheng_Wu:2014}. 
This specific algorithm searches for a curve in the de-shaped spectrogram, whose intensity is maximal in the de-shaped spectrogram, and such that the identified curve is smooth. The smoothness of the curve is controlled by a penalty parameter $\lambda>0$. 
The overall performance is relatively stable over a wide range of $\lambda$, but still there is the need to tune the value of $\lambda$. Furthermore, this approach is not optimal from a computational viewpoint. In this work, we accept that we do not have an extremely well-performing curve extraction method and have to tune the $\lambda$ parameter. Designing a better curve extraction algorithm is by itself an open research problem, and we leave it as a future work.

The whole procedure, including finding the de-shaped spectrogram and extracting curves from the de-shaped spectrogram, form the proposed {\em$\ours$ algorithm}.
%
%

{\subsection{Numerical details of the $\ours$ algorithm}\label{supp:NumericalDetails}

We provide numerical details of the $\ours$ algorithm with all parameters for the reproducibility purpose.

Suppose the PPG signal, denoted as $f$ in the continuous setup, is sampled at a frequency $f_s$ Hz and $N$ points are sampled; that is, we collect the PPG signal for $N/f_s$ seconds. We denote the discretized PPG signal as $\mathbf{f}\in\RR^N$.
To numerically evaluate the STFT defined in (\ref{eq: stft1}), we fix the frequency resolution by $\frac{f_s}{2M}$, where $M\in\NN$, so that the number of discretization points in the frequency axis is $M$ and we take a discretized window function $\mathbf{h}\in \RR^{2k+1}$, $k\in\NN$. The discretization of the STFT of $f$, denoted as $\mathbf{V}_{\mathbf{f}}\in\CC^{N\times2M}$, is thus numerically evaluated by
\begin{equation}
\mathbf{V}_{\mathbf{f}}(n,m) = \frac{1}{f_s}\sum_{l=-k}^{k}\mathbf{f}(n+l) \mathbf{h}(l+k+1)e^{-i\pi n(m-M)/M}\,,
\end{equation}
where $m=1,\ldots,2M$ and we complete $\mathbf{f}$ with $0$'s so that $\mathbf{f}(l)=0$ when $l<1$ and $l>N$. Note that the first $M$ frequency bins correspond to the negative frequency axis while the $M+1,\ldots,2M$-th frequency bins correspond to the non-negative frequency axis.
We fix the quefrency resolution to $\frac{M}{f_sM'}$, where $M'\in\NN$ , so that the number of discretization points in the quefrency axis is $M'$. To numerically implement STCT defined in (\ref{eq: rceps1}), we fix a sufficiently small $\gamma>0$. The discretized STCT, denoted as $\mathbf{C}_{\mathbf{f}}\in\mathbb{C}^{N\times M'}$, is thus obtained as:
\begin{equation}
\mathbf{C}_{\mathbf{f}}(n,m') := \frac{f_s}{2M}\sum_{m=1}^{2M} |\mathbf{V}_{\mathbf{f}}(n,m)|^\gamma e^{-i\pi m' (m-M)/M}, \nonumber
\end{equation}
where $m'=1,\ldots,M'$.

The numerical implementation of the nonlinear mask defined in (\ref{Definition:NonlinearMask}), denoted as $\mathbf{U}_{\mathbf{f}}\in\CC^{N\times M}$, is given by:
\begin{equation}
\mathbf{U}_{\mathbf{f}}(n,m):=
\left\{
\begin{array}{cl}
\displaystyle\sum^{\lceil m+1/2\rceil}_{1/m''=\lceil m-1/2\rceil}\tilde{\mathbf{C}}_{\mathbf{f}}\left(n,\frac{1}{m''}\right)&\mbox{ when }m=1,\ldots,\lceil\frac{f_s}{2M\theta}\rceil\\
&\\
\displaystyle0&\mbox{ otherwise }
\end{array}
\right.
\end{equation}
for all $n=1,\ldots,N$, where $0<\theta<1$ is the threshold chosen by the user for the sake of stabilizing the influence of the low quefrency component in the nonlinear mask design.
$\tilde{\mathbf{C}}_{\mathbf{f}}\in\mathbb{C}^{N\times \alpha M}$ is the upsampled version of $\mathbf{C}_{\mathbf{f}}$ in the quefreqncy axis by $\alpha\in \NN$ times that is defined by
\begin{equation}
\tilde{\mathbf{C}}_{\mathbf{f}}(n,m'') := \mathbf{C}_{\mathbf{f}}\left(n,M+\frac{m''}{\alpha}\right),
\end{equation}
for all $n=1,\ldots,N$ and $m''=1,\ldots,2\alpha M$.

In the nonlinear mask, we only focus on the positive frequency axis, so only $M$ bins corresponding to the non-negative frequency are evaluated.

The de-shape spectrogram of $\mathbf{f}$, denoted as $\mathbf{W}_{\mathbf{f}}\in\mathbb{C}^{N\times M}$, is thus evaluated by discretizing (\ref{definition:SSTW}) as
\begin{equation}
\label{eq:W}
\mathbf{W}_{\mathbf{f}}(n,m) :=
\!\!\!\!\!\!\!\!\!\!\!\!\!\!\!\!\!\!\!\!\!\!\!\!\!\!\!\!\!\!\!\!\!\!\!\!\!\!\!\!\!\!
\sum_{\qquad \qquad \scriptsize k:\left\{\begin{array}{l}
         \!\!\! \scriptsize|\Im\boldsymbol{\Omega}_{\mathbf{f}}(n,k)-m\Delta_\omega|\leq \Delta_\omega/2 \\
         \!\!\! \scriptsize |\mathbf{V}_{\mathbf{f}}(n,k)|\geq\upsilon
        \end{array}
        \right.}
        \!\!\!\!\!\!\!\!\!\!\!\!\!\!\!\!\!\!\!\!\!\!\!\!\!\!\!\!\!\!\!\!\!\!\!\!\!
|\mathbf{V}_{\mathbf{f}}(n,M+k)\mathbf{U}_{\mathbf{f}}(n,k)|,\nonumber
\end{equation}
where $\boldsymbol{\Omega}_{\mathbf{f}}\in \mathbb{C}^{N\times M}$ is the following direct discretization of the reassignment rule (\ref{RM:omega})
\begin{equation}
\boldsymbol{\Omega}_{\mathbf{f}}(n,m):=
\left\{
\begin{array}{ll}
\displaystyle \frac{-i \mathbf{V}_{\mathbf{f}}(n+1,M+m)-\mathbf{V}_{\mathbf{f}}(n,M+m)}{2\pi f_s \mathbf{V}_{\mathbf{f}}(n,M+m)}&\mbox{when }n=1,\ldots,N-1\\
&\\
\displaystyle \frac{-i \mathbf{V}_{\mathbf{f}}(1,M+m)-\mathbf{V}_{\mathbf{f}}(N,M+m)}{2\pi f_s \mathbf{V}_{\mathbf{f}}(N,M+m)}&\mbox{when }n=N\,
\end{array}
\right.
\end{equation}
for $m=1,\ldots,M$.

The curve extraction algorithm is implemented by fitting a discretized curve $\mathbf{c}^*\in Z_M^N$, where $Z_M:=\{1,2,\ldots,M\}$, to the dominant curve in $\mathbf{W}_{\mathbf{f}}$, which is obtained by maximizing the following functional
\begin{equation}
\mathcal{I}(\mathbf{c}):=\sum_{n=1}^N \log\left(\frac{|\mathbf{W}_{\mathbf{f}}(n,\mathbf{c}(n))|}{\sum_{n=1}^N\sum_{m=1}^M|\mathbf{W}_{\mathbf{f}}(n,m)|}\right)-\lambda \sum_{n=2}^N|\mathbf{c}_{n}-\mathbf{c}_{n-1}|^2,\label{Equation:CurveExtraction}
\end{equation}
where $\mathbf{c}\in Z_M^N$ and $\lambda>0$ is the penalty term controlling the regularity of the extracted curve. Clearly, the first term of the functional $\mathcal{I}$ captures the dominant curve in the de-shaped spectrogram.  The final IHR or IRR is obtained by $\frac{f_s}{2M}\mathbf{c}^*$.

In order to have a good accuracy in the extraction of IHR and IRR curves from the PPG signal, we apply a ``divide et impera'' approach. 
We first extract the curve corresponding to the IHR. Then we use it to filter out the cardiac component information from the de-shape spectrogram. We do so by splitting the de-shaped spectrogram of the PPG signal into two parts. The de-shaped spectrogram of the respiratory part of the signal is then analyzed to extract the curve corresponding to the IRR. Regarding the extraction of curves corresponding to the IHR and IRR from the de-shaped spectrogram, we point out that while there exist several good algorithms, how to design an accurate and fully automatic curve extraction algorithm is in general an open problem. In the present work we do not pursue this research problem, but simply apply existing curve extraction algorithms to simplify the discussion. More can be done in this direction in order to improve the performance of the proposed algorithm.

For a fair comparison and for reproducibility purposes, the parameters for computing the de-shaped spectrogram are fixed below. For the ICASSP 2015 signal processing cup database: $f_s=125$ and $M=6,250$, the window size is chosen to be $k=750$ in the STFT, $\gamma=0.3$ in the STCT, $\alpha=5$ and $\theta=0$ in the nonlinear mask design, $\upsilon=10^{-9}$\% of the root mean square energy of the signal under analysis, and $g$ is a direct discretization of the Dirac measure in the de-shaped spectrogram.
For the Capnobase benchmark database: $f_s=300$, $M=30000$, $k=600$, $\gamma=0.3$, $\alpha=5$, $\theta=0$, $\upsilon=10^{-9}$\% of the root mean square energy of the signal under analysis, and $g$ is a direct discretization of the Dirac measure in the de-shaped spectrogram. We mention that the result is not sensitive to the chosen parameters, and no global parameter optimization is carried out in this study, with the only exception for the parameter $\lambda$ used in the curve extraction algorithm, and all parameters are chosen in an ad-hoc matter. The window length is chosen under the empirical rule of thumb that the window should cover few cycles to accumulate enough information, and it should not be too long otherwise the instantaneous information is lost. In practice, choosing a window length from 4 to 10 cycles gives good results.

%

%


}

\section{Results}\label{Section:Results}

In this section, we report the analysis results of the $\ours$ algorithm when applied to the Capnobase benchmark database and the ICASSP 2015 signal processing cup database.
The Capnobase benchmark database (\url{http://www.capnobase.org}) consists of recordings of spontaneous or controlled breathing in static patients. The ICASSP 2015 signal processing cup (\url{http://www.zhilinzhang.com/spcup2015/}), by contrast, contains recordings from
pulse oximeter positioned on the wrists of running subjects.

\subsection{Quantities for evaluation}

\subsubsection{Instantaneous v.s. averaged}

It is necessary to clarify some commonly abused terminologies before reporting the results.
It is well known that ``how fast the heart beats'' and ``how fast one breathes'' provide important physiological information. We commonly use the terminologies, heart rate (HR) and respiratory rate (RR), to refer to them. However, in general, the HR and RR are not scientifically well-defined if the measurement scale is not specified. When the measurement scale is ``infinitesimal'', the HR and RR become IHR and IRR. These instantaneous quantities are what we look for for the physiological variability analysis. 
{When the measurement scale is over seconds or minutes, the HR and RR are derived by counting how many beats or breaths take place over a provided window of length on the scale of seconds or even minutes, of by finding the peak of the power spectrum of the signal over that window. We call the resulting quantities the averaged HR (AHR) and averaged RR (ARR). Clearly, when we mention AHR and ARR, the measurement scale is needed. The commonly encountered quantities regarding ``how frequently the heart beats and how frequently one breathes'' in the PPG literature are AHR and ARR. Note that the IHR and IRR are continuous functions, and the AHR and ARR could be viewed as the low-pass filtered IHR and IRR that comes from a window smoothing technique.

In practice, however, due to the sampling issue, it is in general not possible to measure any physiological system on the infinitesimal scale. The situation is worse when we deal with the HR, for example. Specifically, the IHR is a quantity inaccessible to researchers, and we could only get its surrogate information from the ECG landmarks, like the R peaks. As is recommended by the Task Force \cite{TaskForce:1996}, we could ``recover'' the underling IHR by the interpolation, if the ECG sampling rate is high enough, and the R peak locations are determined accurately. Based on this recommendation, a practical expectation is having an HR estimate on the scale as small as possible, like $10$ msec or $1$ msec. We will take the IHR obtained from the R-peak to R peak interval (RRI) time series\footnote{If the ECG signal was not recorded at 1000 Hz, resample it to be of 1000Hz. This upsample step could give at least millisecond precision in locating the time of the R-wave peak \cite{Laguna2000}. Then, apply the standard R-peak detection algorithm to extract the R-peaks, and hence the R-peak to R peak interval (RRI) time series determined by two consecutive heart beats.} as the gold standard.

As we will show below, the proposed $\ours$ algorithm allows us to directly compute the IHR and the IRR {\em on the scale of $10$ msec}, and in the sequel, when we mention ``instantaneous'', it is this scale that we refer to. With the IHR and IRR, we can easily derive the averaged quantities AHR and ARR by taking the averaging process.}
%

{

\subsubsection{Continuous v.s. discrete quantities}

While the variations of IHR and IRR are commonly applied to quantify the physiological variability, the other signal that is also commonly analyzed to evaluate the physiological variability is the ``beat-to-beat interval'' time series, for example, the R-peak to R peak interval (RRI) time series. 
Take the IHR and RRI as an example. Note that the IHR is a continuous signal, while the RRI is a time series. The IHR and RRI are intimately related and could be converted from one to the other. Precisely, based on the suggestion from Task Force \cite{TaskForce:1996}, we could derive the IHR from the RRI time series via interpolation. On the other hand, we could recover the RRI time series from the IHR via searching the pre-assigned landmark of the ECG signal. 
In general, we obtain one quantity, and recover the other one via the above-mentioned conversion.

For the ECG signal, the conversion between the IHR and the RRI time series seems transparent, and it is easier to find the RRI time series first by the standard R peak detection algorithm. We might extrapolate this fact and expect that finding the peaks is easier than finding the instantaneous frequency in other oscillatory signals.
However, we claim that it is not the case. Instead, for most oscillatory signals, finding the instantaneous frequency could be easier and more stable than finding the peaks. 

Notice that the main mechanism of the R peak detection algorithm is based on the easy-to-define and easy-to-identify landmark in the ECG signal, like the R peak. However, for most oscillatory signals, such landmark is not easy to find, even to define. Take the respiratory oscillation in the PPG signal into account. It is clear that while there is an oscillation, it is not easy to define a reliable landmark in the PPG signal for the respiratory component so that we could determine the ``beat-to-beat interval'' time series for the respiration. 
Another scenario in which the peak detection is not favorable is when noise exists and the landmark is not ``stable''. The R peak could be view as a stable landmark in the sense that it is robust to noise. Often, the ``peak'' of most oscillatory signals, even if it is possible to provide a universally accepted definition for ``peak'', is not stable when noise exists. For example, the respiratory signal recorded from the capnogram. 

In summary, depending on the type of oscillatory signal, we might favor to directly estimate the continuous quantity. In this paper, we focus on the continuous quantity and report the comparison.

\subsection{Simulated database}\label{Section:SimulationSemi}

To validate the proposed $\ours$ algorithm in extracting the instantaneous physiological dynamics from the PPG signal, we consider the following semi-real simulated PPG database. First, we take a 5-min PPG signal sampled at $300$ Hz, recorded from Philip patient monitor MP60, from $5$ normal subjects under general anesthesia. The 5-min period is chosen to contain the endotracheal intubation and other manipulation so that the PPG signal we will simulate contains enough dynamical changes. Truncate the signal into pieces according to the oscillations, where the oscillations are visually confirmed. Then, align all cycles so that their peaks are aligned, and trim each cycles so that they have the same length.
Denote the trimmed cycles $\mathbf P^{(0)}_i\in \mathbb{R}^p$, where $i=1,\ldots, N_0$ and $p,N_0\in \mathbb{N}$. We remove the outliers in the following way. Evaluate a median cycle out of all cycles $\mathbf P^{(m)}\in \mathbb{R}^p$ so that $\mathbf P^{(m)}(j)=\text{median}\{\mathbf P^{(0)}_i(j)\}_{i=1}^{N_0}$, where $j=1,\ldots,p$. Then collect all cycles that have the $L^2$ distances from $\mathbf P^{(m)}$ within the $90\%$ quantile of $\{\|\mathbf P^{(m)}-\mathbf P^{(0)}_i\|\}_{i=1}^{N_0}$. Suppose there are $K<N_0$ cycles left. To avoid the boundary effect, we subtract each cycle by a linear function so that the resulting cycle starts and ends at $0$. Finally, collect those post-processed cycles as $\mathcal{C}:=\{\mathbf P_i\}_{i=1}^K$, and view each $\mathbf P_i$ as a periodic function. 
We call $\mathcal{C}$ a {\em standardized PPG cycle set}. Second, for each subject in the Capnobase benchmark database, evaluate the IHR derived from the RRI of the ECG signal by the cubic spline interpolation according to the Task Force guidance \cite{TaskForce:1996}. Since the purpose of this simulated database is to validate the proposed algorithm in getting the instantaneous physiological information, to avoid the influence coming from the bad ECG signal quality, we focus on the IHR signal over a continuous $3$ minutes period where the ECG signal is not contaminated by big noise. Note that the possibly existing arrhythmia is not taken into account when evaluating the signal quality. 
Denote the selected 3-min IHR as $\textsf{IHR}_{\textsf{ECG}}(t)$, and the phase of the IHR as $\phi_{\textsf{ECG}}(t)$; that is, $\frac{d}{dt}\phi_{\textsf{ECG}}(t)=\textsf{IHR}_{\textsf{ECG}}(t)$. Then, sample $\phi_{\textsf{ECG}}$ at $300$Hz and denote the discretized phase function as $\mathbf{\phi}\in\mathbb{R}^n$. Third, simulate a semi-real PPG signal from a standardized PPG cycle set $\mathcal{C}$ and a phase function $\mathbf{\phi}$ by the following stitching step. Denote $\mathbf{Q}$ to be the modulus of $\mathbf{\phi}$ after being divided by $1$, which is clearly a sawtooth function. Denote $\{\iota_j\}_{i=1}^I\subset \mathbb{N}$ so that $\iota_j$ is the $j$-th jump of $\mathbf{Q}$ from $1$ to $0$. The hemodynamic component of the simulated semi-real PPG signal, denoted as $\mathbf h\in\mathbb{R}^n$, is constructed by
\begin{equation}
\mathbf h(i)=\left\{
\begin{array}{ll}
\mathbf P_{\tilde k}(\textsf{round}(\mathbf{\phi}(i)))& \mbox{when }i =\iota_k+1\ldots,\iota_{k+1}\\
0& \mbox{otherwise},
\end{array}
\right.
\end{equation}
where $\textsf{round}(x)$ rounds the real number $x$ to the nearest integer, $i=1,\ldots,n$, and $\tilde k$ is the modulus of $k$ after being divided by $K$. The respiratory component of the simulated semi-real PPG signal, denoted by $\mathbf r\in\mathbb{R}^n$, is simply taken from the capnometry signal after scaling it so that its mean is a value randomly and uniformly chosen from the interval between one quarter and half of the mean of $\mathbf  h$. The clean semi-real PPG signal is constructed by $\mathbf f_0:=\mathbf h+ \mathbf r $. To model the inevitable noise, we consider an autoregressive and moving average $(1,1)$ time series $\xi\in \mathbb{R}^n$, which is determined by the autoregression polynomial $a(z)=0.95z-1$ and the moving averaging polynomial $b(z)=0.5z+1$, with the innovation process taken as independently and identically distributed Student $t_4$ random variables. We model the noise in order to capture the possibly time-dependent structure in the noise and the spike-like artifacts commonly seen in the real data.
The final simulated semi-real PPG signal is $\mathbf f=\mathbf f_0 +\sigma\xi\in \mathbb{R}^n$, where $\sigma>0$ is the strength of the noise, which is chosen to satisfy the desired signal-to-noise ratio (SNR), where the SNR is defined as $20\log_{10}\frac{\text{std}(\mathbf f_0)}{\text{std}(\sigma\xi)}$, and the unit is dB.

For each simulated PPG signal, we denote the estimated IHR from $\ours$ as $\textsf{IHR}_{\ours}(t)$, and evaluate the RMS of estimating $\textsf{IHR}_{\textsf{ECG}}(t)$ by $\textsf{IHR}_{\ours}(t)$ to evaluate the algorithm's performance. 

Among those 42 subjects in the Capnobase, the signal qualities of \texttt{0030\textunderscore 8min}, \texttt{0115\textunderscore 8min}, \texttt{0147\textunderscore 8min}, \texttt{0309\textunderscore 8min} and \texttt{0329\textunderscore 8min} are not ideal, so the IHR's are selected from a 3-min interval with a reasonable ECG quality. For the other subjects, we take the first 3 mins IHR for the simulation. With 4 subjects under general anesthesia, we obtain 4 standardized PPG cycle sets. As a result, we could realize $168$ semi-real simulated PPG signals for each SNR. 

See Figure \ref{fig:Simulation} for an illustration of one typical realization of the simulation with the $20$dB SNR and the analysis result. 
It is clear that the spectrogram is broadened, which is caused by the uncertainty principal shared by all linear-type time-frequency analysis tools. It is not easy to capture the ``instantaneous'' information about the hemodynamics from this broadened spectrum (the red arrow). Furthermore, the multiples of the fundamental frequency of the hemodynamic component are clearly seem in the spectrogram, which makes it difficult to interpret (the green arrows). Even worse, the respiratory component in the PPG signal contributes to the low frequency region (the blue arrow), while its multiples are relatively weak and cannot be clearly seen in the figure. Such limitations are alleviated directly after applying this de-shape nonlinear masking technique. In the de-shape spectrum, only a clear, dominant, and concentrated line exists, which represents the instantaneous hemodynamic information. In this case, the RMS is $13.51$ msec. 

When the SNR is $20$ dB, the RMS is $16.02\pm 7.79$ msec, where the results are represented as ``mean $\pm$ standard deviation'' over $168$ realizations. Overall, we could obtain the heart rate information in the instantaneous sense on the scale of around $20$ msec. 
In practice, the noise sometimes could be large. Consider the case when the SNR is $10$ dB. In this case, the RMS is $20.94\pm 8.07$ msec. It is clear that when the noise is larger, the algorithm's performance is worse, but the performance degradation is stable to the noise.

\begin{figure}
\centering
\includegraphics[width=0.9\columnwidth]{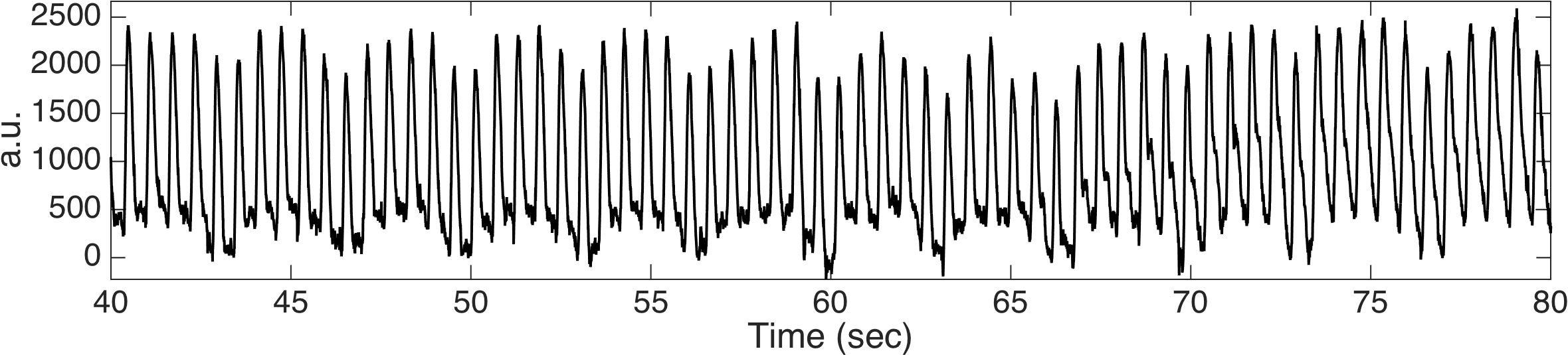}
\includegraphics[width=0.9\columnwidth]{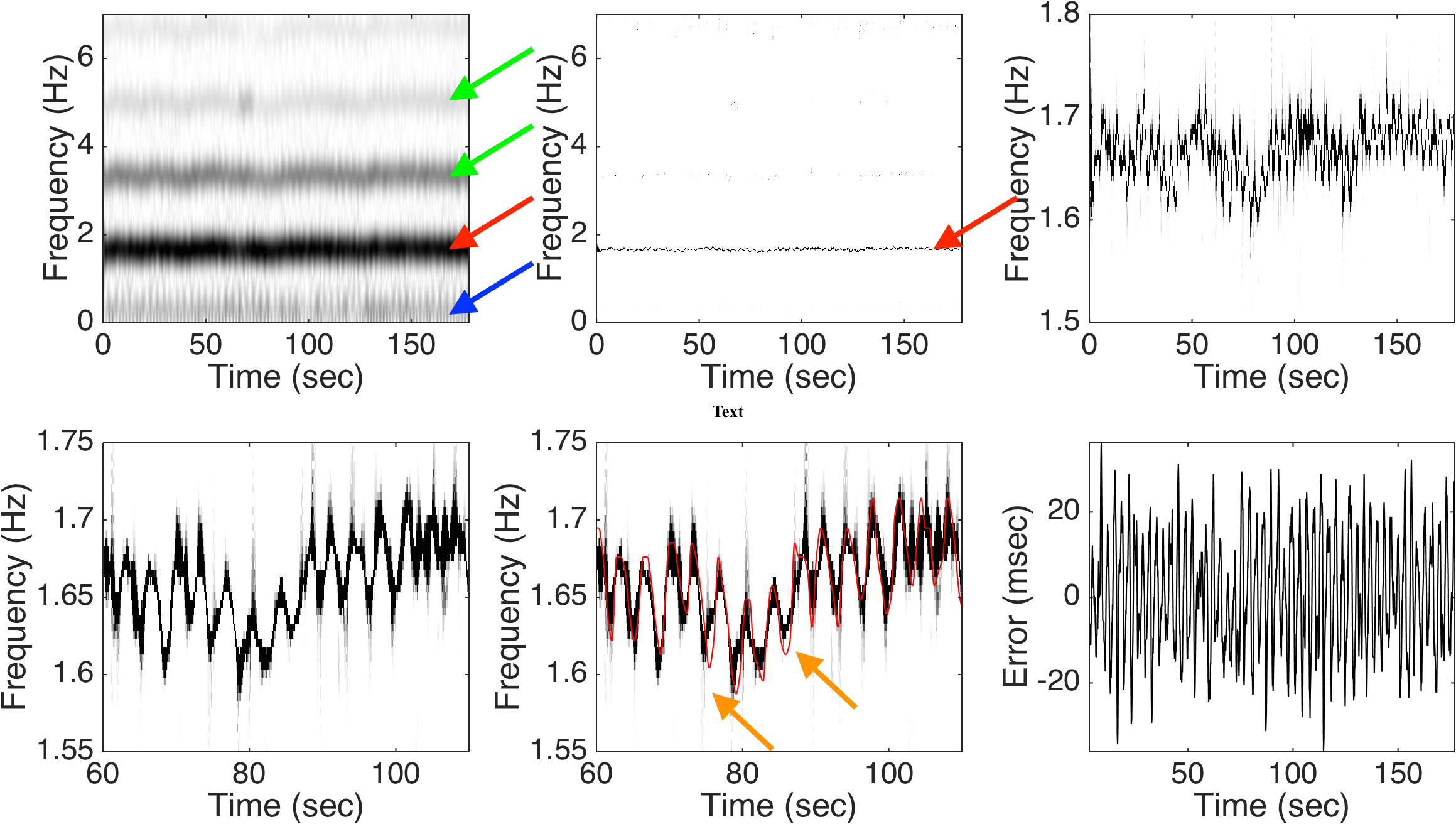}
\caption{{Top panel: a subinterval of the simulated photoplethysmogram (PPG) signal with the instantaneous heart rate (IHR) determined from \texttt{0009\textunderscore 8min} in the Capnobase database and $20$dB signal-to-noise ratio. a.u. means ``arbitrary unit''. Middle panel: the spectrogram determined by the short time Fourier transform (STFT) is shown on the left hand side, the de-shaped spectrogram is shown in the middle, and the zoomed in de-shaped spectrogram is shown on the right hand side. The red arrow indicates the band associated with the fundamental frequency of the hemodynamic component, the green arrows indicate those bands associated with the multiples of the hemodynamic component, and the blue arrow indicates the band associated with the fundamental frequency of the respiratory component. Bottom panel: a further zoomed in de-shape spectrogram is shown on the left hand side and the middle, with the ground truth IHR superimposed in the middle, and the error between the estimated IHR and the ground truth IHR is shown on the right hand side. The orange arrows indicate the possible oversmoothing effect of the $\ours$ algorithm. See the main context for the interpretation of this figure.}}
\label{fig:Simulation}
\end{figure}
}

\subsection{Capnobase benchmark database -- PPG signal with respiration}

The Capnobase benchmark database includes forty-two eight minute segments from 29 pediatric and 13 adults cases containing reliable recordings of spontaneous or controlled breathing. For each subject, the ECG, capnometry, and PPG signals were recorded at 300 Hz, 25 Hz, and 100 Hz respectively. All signals were recorded with S/5 Collect software (Datex--Ohmeda, Finland). The PPG and capnometry were automatically up-sampled to have a 300 Hz sampling rate. The database also contains a reference ARR, as well as information regarding the beginning of each expiration, both derived from the capnogram waveform and identified by experts. The reference AHR as well as R-peak locations derived by experts from the ECG waveform are also provided.

The $\ours$ method provides estimates for the IHR and IRR, which are instantaneous in nature. However, it is important to remember that, to the best of our knowledge, the methods proposed so far in the literature for the Capnobase benchmark datasets provide AHR and ARR information. They do not focus on computing \textit{instantaneous rates}, but \textit{average rates} over a time window. {In most cases, the window is set to be around 60 seconds; that is, they report the HR on the scale of 60 seconds.}
Thus, for a fair comparison, we also provide AHR and ARR curves obtained from the estimated IRR and IHR by smoothing over a 60 seconds window shifted of 30 seconds each time. We refer to this variation of the method as \emph{$\ours$-60s}.

We denote $x^{\texttt{ref}}_1,\ldots,x^{\texttt{ref}}_n$, where $n$ is the number of observations, to be the reference information, i.e. IHR or IRR (respectively AHR or ARR) which are either provided directly by the experts and included in the database, or derived from the R peaks and the beginning timestamps of expiration respectively. We denote $x^{\texttt{alg}}_1,\ldots,x^{\texttt{alg}}_n$ to be the estimated IHR or IRR determined by $\ours$ (respectively the estimated AHR or ARR determined by $\ours$-60s). Following what has been done in the literature, we assess the performance of the proposed algorithm using the RMS and the mean absolute error (MAE), which are defined as
\begin{equation}\label{eq:RMS_error}
   \textrm{MAE} = \frac{1}{n}\sum_{i=1}^n\left|x_i^{\texttt{ref}}-x_i^{\texttt{alg}}\right|
\end{equation}

The challenge of simultaneously estimating IHR and IRR from a single PPG signal is made more difficult by the possible presence of artifacts.  The Capnobase benchmark database include in each dataset information regarding potential intervals in the PPG, ECG and Capnometry waveforms that contain artifacts. The methods reported in the literature use this information to skip the measurement errors over windows, if not even entire datasets, that are considered unreliable; see, for example, \cite{Dehkordi2015}.
In our tests, we evaluate the performance of the proposed $\ours$ method using all of the 42 datasets, without excluding any of their intervals, even those that are supposed to contain artifacts. We do so to test the performance of the $\ours$ algorithm in real scenarios.
It is clear that for the intervals labeled as containing artifacts, the ground truth provided in the database is simply given as an interpolation of the nearby reliable values; see the right panel in Figure SI.2. 
This introduces a bias in the error values we compute. In particular, the performance of the proposed method can be consistently improved, if we remove intervals containing artifacts. However, in a real life application, such information is clearly not available in general. We discuss this in more detail in Section C in the online SI.

The RMS and MAE of the $\ours$ and $\ours$-60s are then evaluated and reported
in Tables \ref{tab:Capno_stats}. These tables provide the summary statistics of the RMS and MAE for the respiratory and heart rate obtained from the proposed algorithm. The performance of other methods proposed so far in the literature, and their chosen windows for averaging, are also included for a comparison.
From these tables, it is clear that the $\ours$ provides a satisfactory IHR and IRR estimation, and the AHR and ARR provided by $\ours$-60s perform better than the other methods proposed in the literature. {Specifically, the RMS of the IHR by $\ours$ is $0.93$ beats per minute, which is equivalent to $15.5$ msec. This result could be interpreted as obtaining the HR in the instantaneous sense on the scale of around $15$ msec. A similar interpretation holds for the IRR result. }

An example illustrating how the $\ours$ analyzes the PPG signal of the dataset \texttt{0009\textunderscore 8min} in the Capnobase benchmark database, is shown in the algorithm flowchart in Figure \ref{FlowChart}. By a visual inspection, it is clear that there are two oscillations inside the PPG signal -- the faster (respectively slower) oscillations are associated with the heartbeat (respectively respiration). It is clear that the spectrogram of the PPG signal is complicated by the interfering multiples of these two oscillations, which are due to their non-sinusoidal shape as explained in the previous section. However, this interference is eliminated in the de-shaped spectrogram.
%

\begin{table}[H]
\begin{minipage}{.8\textwidth}
{\footnotesize
\begin{tabular}{|c||c|c|c|c|c||c|c|c|c|c|}\hline
  \multirow{2}{*}{RR (breaths/minute)} & \multicolumn{5}{c||}{RMS} & \multicolumn{5}{c|}{MAE} \\ \cline{2-11} & mean & std & $Q_1$ & median & $Q_3$  & mean & std & $Q_1$ & median & $Q_3$ \\ \hline
\hline
Smart Fusion \cite{Karlen2013}\footnote{\label{Karlen2013}Sliding window of 32s with 1s shifts.} & 3.00 & 4.70 & 0.60 & 1.56 & 3.15 & 2.43 & 3.72 & {\tiny n/a} & {\tiny n/a} & {\tiny n/a}\\ \hline
CSD - 120s \cite{Garde2014}\footnote{\label{Garde2014_120CSD}Sliding window of 120s with 50\% overlap. CSD: correntropy spectral density. PSD: power spectral density.} & {\tiny n/a} & {\tiny n/a} & 0.27 & 0.95 & 6.20 & {\tiny n/a} & {\tiny n/a} & {\tiny n/a} & {\tiny n/a} & {\tiny n/a}\\ \hline
CSD - 60s \cite{Garde2014}\footnote{\label{Garde2014_60CSD}Sliding window of 60s with 50\% overlap.} & {\tiny n/a} & {\tiny n/a} & {\tiny n/a} & 1.77 & {\tiny n/a} & {\tiny n/a} & {\tiny n/a} & {\tiny n/a} & {\tiny n/a} & {\tiny n/a}\\ \hline
PSD - 120s \cite{Garde2014}\footref{Garde2014_120CSD} & {\tiny n/a} & {\tiny n/a} & 1.20 & 3.18 & 11.30 & {\tiny n/a} & {\tiny n/a} & {\tiny n/a} & {\tiny n/a} & {\tiny n/a}\\ \hline
Garde \textit{et al.} \cite{garde2013empirical}\footref{Garde2014_60CSD} & {\tiny n/a} & {\tiny n/a} & 1.10 & 3.50 & 11.00 & {\tiny n/a} & {\tiny n/a} & {\tiny n/a} & {\tiny n/a} & {\tiny n/a}\\ \hline
Shelley \textit{et al.} \cite{Shelley2006}\footnote{Sliding window of 82s.} & {\tiny n/a} & {\tiny n/a} & 0.41 & 1.91 & 7.01 & {\tiny n/a} & {\tiny n/a} & {\tiny n/a} & {\tiny n/a} & {\tiny n/a}\\ \hline
Nakajima \textit{et al.} \cite{Nakajima1996}\footnote{No information is provided on the window size.} & {\tiny n/a} & {\tiny n/a} & 0.59 & 7.47 & 10.60 & {\tiny n/a} & {\tiny n/a} & {\tiny n/a} & {\tiny n/a} & {\tiny n/a}\\ \hline
Zhang et al. \cite{Zhang2016}\footnote{\label{Zhang2016}Sliding window of 32s with an increment of 3s. BCLA: Bayesian Continuous-Valued Label Aggregator.} & 2.44 & 4.08 & {\tiny n/a} & {\tiny n/a} & {\tiny n/a} & 1.52 & 2.73 & {\tiny n/a} & {\tiny n/a} & {\tiny n/a}\\ \hline
BCLA - Zhu \textit{et al.} \cite{Zhu_Pimentel:2015}\footref{Zhang2016} & {\tiny n/a} & {\tiny n/a} & {\tiny n/a} & {\tiny n/a} & {\tiny n/a} & 1.97 & 0.40 & {\tiny n/a} & {\tiny n/a} & {\tiny n/a}\\ \hline
Dehkordi \textit{et al.} \cite{Dehkordi2015}\footnote{The values reported are obtained excluding 4 datasets from the statistics
due to contamination of their PPG or CO2 signals with artifacts for more than 50\% of their duration. However, the authors do not specify which datasets have been removed. No information is provided regarding the window length.} & {\tiny n/a} & {\tiny n/a} & {\tiny n/a} & 0.39 & {\tiny n/a} & {\tiny n/a} & {\tiny n/a} & {\tiny n/a} & {\tiny n/a} & {\tiny n/a}\\ \hline
$\ours$ vs reference IRR & 1.39 & 1.87 & 0.38 & 0.73 & 1.70 & 0.94 & 1.37 & 0.22 & 0.50 & 0.83\\ \hline
$\ours$-60s vs reference ARR & 0.78 & 1.60 & 0.09 & 0.22 & 0.62 & 0.53 & 1.16 & 0.07 & 0.15 & 0.43\\ \hline
\hline
\multirow{2}{*}{HR (beats/minute)} & \multicolumn{5}{c||}{RMS} & \multicolumn{5}{c|}{MAE} \\ \cline{2-11} & mean & std & $Q_1$ & median & $Q_3$  & mean & std & $Q_1$ & median & $Q_3$ \\ \hline
\hline
Smart Fusion \cite{Karlen2013}\footref{Karlen2013} & {\tiny n/a} & {\tiny n/a} & 0.37 & 0.48 & 0.77 & {\tiny n/a} & {\tiny n/a} & {\tiny n/a} & {\tiny n/a} & {\tiny n/a}\\ \hline
CSD - 120s \cite{Garde2014}\footref{Garde2014_120CSD} & {\tiny n/a} & {\tiny n/a} & 0.34 & 0.76 & 1.45 & {\tiny n/a} & {\tiny n/a} & {\tiny n/a} & {\tiny n/a} & {\tiny n/a}\\ \hline
PSD - 120s \cite{Garde2014}\footref{Garde2014_120CSD} & {\tiny n/a} & {\tiny n/a} & 0.21 & 0.58 & 1.17 & {\tiny n/a} & {\tiny n/a} & {\tiny n/a} & {\tiny n/a} & {\tiny n/a}\\ \hline
Garde \textit{et al.} \cite{garde2013empirical}\footref{Garde2014_60CSD} & {\tiny n/a} & {\tiny n/a} & 0.20 & 0.35 & 0.59 & {\tiny n/a} & {\tiny n/a} & {\tiny n/a} & {\tiny n/a} & {\tiny n/a}\\ \hline
Wadehn \textit{et al.} \cite{Wadehn2015}\footnote{\label{8-6_Capnobase}Sliding window of 8s with 6s overlap.} & {\tiny n/a} & {\tiny n/a} & {\tiny n/a} & {\tiny n/a} & {\tiny n/a} & 0.16 & 0.24 & {\tiny n/a} & {\tiny n/a} & {\tiny n/a}\\ \hline
$\ours$  vs reference IHR & 0.93 & 0.57 & 0.50 & 0.72 & 1.20 & 0.61 & 0.35 & 0.38 & 0.52 & 0.84\\ \hline
$\ours$-60s  vs reference AHR & 0.23 & 0.49 & 0.07 & 0.09 & 0.19 & 0.15 & 0.29 & 0.05 & 0.07 & 0.12\\ \hline
\end{tabular}
}
\caption{Summary of root mean square error (RMS) and mean absolute error (MAE) of the respiratory rate (RR) and heart rate (HR) estimation for the Capnobase benchmark database. The unit for the RR is breaths per minute, and the unit for the HR is beats per minute. Except $\ours$, the methods proposed so far in the literature do not focus on computing \textit{instantaneous rates}, but \textit{average rates} over a time window. n/a: not available. Std: standard deviation. $Q_1$: first quartile. $Q_3$: third quartile.}
\label{tab:Capno_stats}
\end{minipage}
\end{table}

\subsection{ICASSP 2015 signal processing cup -- PPG signal with motion}

The second database we consider is the training database of the ICASSP 2015 signal processing cup, which contains recordings of 12 male subjects with yellow skin and ages ranging from 18 to 35. Two-channel PPG signals, three-axis acceleration signals and a one-channel ECG signal were simultaneously recorded from the subjects. For each subject, the PPG signals were recorded from the dorsal of the wrist by two pulse oximeters with green LEDs (wavelength: 515nm), and the distance between two PPG sensors (from center to center) was 1.5 cm \cite[and private communication]{Zhang2015}.

Each dataset is a 5-minute PPG signal sampled at 125Hz and recorded when a subject runs on a treadmill with changing speeds, scheduled as follows: for subject one, the schedule is rest (30s) $\rightarrow$ 8km/h (1min) $\rightarrow$ 15km/h (1min) $\rightarrow$ 8km/h (1min) $\rightarrow$ 15km/h (1min) $\rightarrow$ rest (30s); for
subject 2 to subject 12, the schedule is rest (30s) $\rightarrow$ 6km/h (1min) $\rightarrow$ 12km/h (1min) $\rightarrow$ 6km/h (1min) $\rightarrow$ 12km/h (1min) $\rightarrow$ rest (30s).
The subjects were asked to purposely use the hand with the wristband to pull clothes, wipe sweat on the forehead, and push buttons while they run on the treadmill, in addition to freely swing.
The acceleration signal was also recorded from the wrist using a three-axis accelerometer. Both the pulse oximeter and the accelerometer were embedded in a wristband, which was comfortably worn. The ECG signal was recorded from the chest using wet ECG sensors. Based on the ECGs, the reference AHR over an 8-second time window evaluated by experts is provided in the database.

To evaluate the performance of the proposed $\ours$ algorithm, and to have a fair comparison with other reported results, for this database we follow the available literature and consider the average absolute error (AAE), which is the same as MAE defined in (\ref{eq:RMS_error}), and the average absolute error percentage (AAEP):
\begin{equation}\label{eq:RMS_error}
    \textrm{AAEP} = \frac{1}{n}\sum_{i=1}^n\frac{\big|x_i^{\texttt{ref}}-x_i^{\texttt{alg}}\big|}{x_i^{\texttt{ref}}}.
\end{equation}
The summary of the HR estimation by different methods proposed in the literature and our method is shown in Table \ref{tab:Motion_HR_AAEP}. Note that the result by $\ours$ is consistently better than the existing reported results.
Since we do not have the ground truth information about the RR, we do not evaluate it for this database.

\begin{figure}[ht]
\centering
\includegraphics[width=0.9\linewidth]{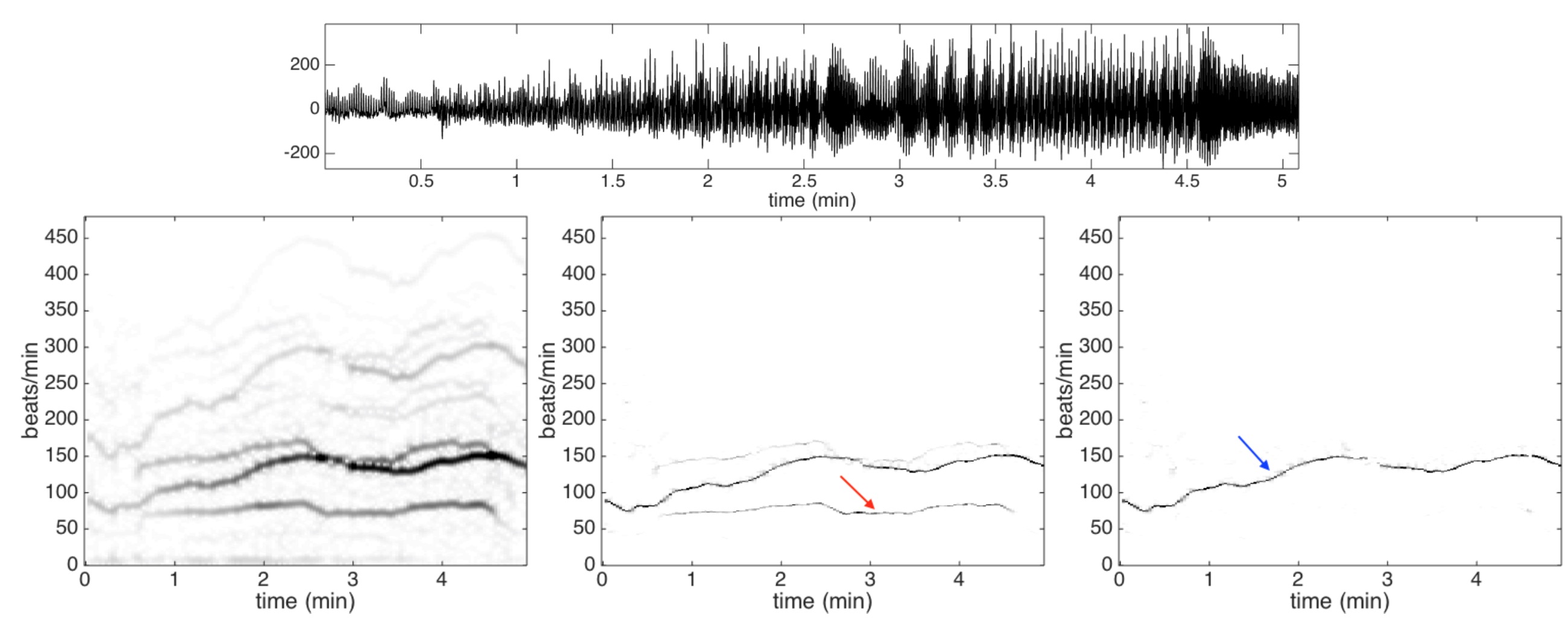}
\caption{Top row: the photoplethysmography signal of subject 9 in the training dataset of ICASSP 2015 signal processing cup. Second row: the spectrogram is shown on the left panel, the de-shaped spectrogram is shown in the middle panel, and the de-shaped spectrogram with the acceleration offset is shown on the right panel. The dominant curve indicated by the blue arrow in the de-shaped spectrogram with the acceleration offset is the instantaneous heart rate of the subject. On the other hand, the lighter curve indicated by the red arrow in the de-shaped spectrogram is directly related to the body swaying pattern. The heartbeat component displayed in the de-shaped spectrogram follow clearly the running pattern.}\label{fig:MotionPPG_TFR}
\end{figure}

An important factor to be taken into account is that PPG signals are prone to motion artifacts, even in presence of small motion \cite{Zhang2014}. The PPG signals under analysis have been sampled during intensive physical exercise. Therefore, we expect to have strong motion artifacts.
To see the contribution of the motion artifacts that do not exist in the Capnobase benchmark database, we show in Figure \ref{fig:MotionPPG_TFR} the PPG signal of subject 9 of the training dataset. In the same figure, we show the spectrogram of that PPG signal as well as the de-shaped spectrogram.
From the recorded signal, which is shown in the top row of Figure \ref{fig:MotionPPG_TFR}, it is not easy to see how the motion and heartbeat vary. The IHR displayed in the de-shaped spectrogram in Figure \ref{fig:MotionPPG_TFR} clearly follow the running pattern. It starts around 90 beats per minute and it goes up to roughly 150 beats per minute after 2.5 minutes, then it goes down for a minute and then again up for another one.

Note that the IHR lies between two other components supposedly affected by motion artifacts, which can be observed in the left and central panels in the bottom row of Figure \ref{fig:MotionPPG_TFR}. The higher frequency component associated with motion has instantaneous frequency about twice the instantaneous frequency of the lower one.
We conjecture that the higher one is contributed by the movement of body while the lower is contributed by the movement of arms and legs. The body finishes a period by just one step, while the leg finishes a period by two steps (one leg needs to finish a forward and backward movement). It is quite natural to catch two components here as they are indeed (at least) two different oscillatory signals, where one has an instantaneous frequency almost twice that of the other one.

In order to remove the contribution of the motion artifacts from a PPG signal, we use the acceleration signal that is recorded using a three-axis accelerometer embedded in a wristband. The spectrogram of the acceleration signal is evaluated and used to mask out the motion artifacts in the de-shaped spectrogram.
An example of the final de-shaped spectrogram is shown in Figure \ref{fig:MotionPPG_TFR}.

Another observation regards how the reference AHR is provided in the database. Such a reference curve is computed by counting the number of heart beats over the time window eight seconds in length, thus we could view the reference AHR as the IHR curve smoothed over a window eight seconds in length. However, since the $\ours$ algorithm calculates the IHR directly, there are inevitable ``high frequency components'' in the extracted curves compared with the reference AHR.
Thus, in order to have a fair comparison, we filter out the high frequency components contained in such curves.

This can be done in many different ways. We use a modern signal filtering algorithm called the \textit{Iterative Filtering} that allows us to automatically detect and remove high frequency components from a non-stationary and non-linear signal \cite{cicone2016adaptive,cicone2016hyperspectral,cicone2017MIF}. We call this variation the $\ours$-IF algorithm.
More details about the analysis results are available in Section C.2 in the SI.

In order to measure performance of the proposed $\ours$ method using instantaneous values, we compute a new {\em reference IHR} from the R-peaks of the ECG signals provided in the database. In particular, for every instant of time in the middle of two consecutive R-peaks in the ECG signal, we compute the reciprocal of the time period between these two R-peaks. Based on these values, and following the standard protocol \cite{TaskForce:1996}, we derive the reference IHR curve as a cubic spline interpolation. We compare these curves with the ones produced by the $\ours$ algorithm. The performance are shown in Table \ref{tab:Motion_HR_AAEP}. {Note that the AAE (or RMS) of the IHR by $\ours$ is $2.97$ beats per minute, which is equivalent to $49.5$ msec, which means that we obtained the IHR on the scale of around $50$ msec. While at the first glance this result is worse than that of the Capnobase, note that the subjects in this database are exercising, so the larger error is expected.}

In the online SI, we also report in Table SI.3 the performance obtained using data downsampled to 25 Hz from the original 125 Hz, as well as truncated as proposed in \cite{Zhang2015}.

\begin{table}[H]
\begin{minipage}{\textwidth}
{\tiny
\begin{tabular}{|c|c|c|c|c|c|c|c|c|c|c|c|c||c|c|}\hline
{\tiny AAE (beats/minute)} & {\tiny Sbj 1} & {\tiny Sbj 2} & {\tiny Sbj 3} & {\tiny Sbj 4} & {\tiny Sbj 5} & {\tiny Sbj 6} & {\tiny Sbj 7} & {\tiny Sbj 8} & {\tiny Sbj 9} & {\tiny Sbj 10} & {\tiny Sbj 11} & {\tiny Sbj 12} & {\tiny mean} & {\tiny std}  \\ \hline
\hline
{\tiny TROIKA - Zhang \textit{et al.} \cite{Zhang2014}}\footnote{\label{8-6}8s windows with 6s overlap} & 2.29 & 2.19 & 2.00 & 2.15 & 2.01 & 2.76 & 1.67 & 1.93 & 1.86 & 4.70 & 1.72 & 2.84 & 2.34 & 0.83\\ \hline
{\tiny Schack \textit{et al.} \cite{Schack2015}}\footnote{\label{Schack2015}8s windows. The overlap is unknown}& 2.40 & 1.21 & 1.20 & 1.22 & 1.34 & 1.44 & 1.16 & 1.04 & 1.18 & 5.33 & 2.18 & 1.52 & 1.77 & 1.20\\ \hline
{\tiny SPECTRAP - Sun \textit{et al.}} \cite{Sun2015}\footref{8-6}\footnote{\label{Hz_U}The authors compare their results with the ones of \cite{Zhang2015} where signals are downsampled to 25 Hz. However the same authors mention that the signals are sampled at 125 Hz and no references to downsampling are given.}& 1.18 & 2.42 & 0.86 & 1.38 & 0.92 & 1.37 & 1.53 & 0.64 & 0.60 & 3.65 & 0.92 & 1.25 & 1.39\footnote{\label{WVM}This value has been recomputed using Matlab \textit{mean} function and does not match the value reported in the original paper.} & 0.86\footnote{\label{BHO}This value has been recomputed using Matlab \textit{std} function and does not match the value reported in the original paper.}\\ \hline
{\tiny MICROST - Zhu \textit{et al.}} \cite{Zhu_Tan:2015}\footref{8-6}& 2.93 & 3.06 & 2.03 & 2.29 & 2.64 & 2.58 & 1.97 & 1.77 & 1.87 & 3.81 & 1.91 & 4.07 & 2.58 & 0.77\footref{BHO}\\ \hline
{\tiny MISPT - Murthy \textit{et al.} \cite{murthy2015}}\footref{8-6}& 1.58 & 1.80 & 0.58 & 0.99 & 0.74 & 0.93 & 0.73 & 0.45 & 0.41 & 3.60 & 0.88 & 0.68 & 1.11 & 0.89\footref{BHO}\\ \hline
{\tiny Zong \textit{et al.} \cite{Zong2015}}\footnote{\label{none}The authors did not mention in their work if they used windows or not.}& 1.05 & 0.98 & 1.26 & 1.33 & 0.66 & 0.77 & 0.41 & 0.47 & 0.35 & 3.49 & 0.50 & 1.52 & 1.07 & 0.86\\ \hline
{\tiny Frigo \textit{et al.} \cite{Frigo2015}\footref{8-6}}& 2.11 & 1.89 & 1.01 & 1.08 & 0.61 & 1.66 & 0.54 & 0.59 & 0.54 & 4.12 & 1.15 & 2.83 & 1.51 & 1.09\\ \hline
{\tiny D'souza \textit{et al.} \cite{SandeepJarChakraborti2015}}\footref{8-6}& 3.93 & 3.30 & 2.81 & 2.07 & 0.90 & 2.50 & 0.83 & 1.08 & 0.75 & 3.68 & 1.65 & 3.60 & 2.26 & 1.21\\ \hline
{\tiny Zhang \textit{et al.} \cite{ZhangLiuZhang2015a}}\footref{8-6}& 2.06 & 3.59 & 0.92 & 1.54 & 0.97 & 1.64 & 2.25 & 0.63 & 0.62 & 4.62 & 1.30 & 1.80 & 1.83 & 1.21\\ \hline
{\tiny Mashhadi \textit{et al.} \cite{Mashhadi2015}}\footnote{\label{Mashhadi2015}8s windows. The overlap is unknown. The authors do not specify if they are using truncated or untruncated datasets and if they are sampling at 25 or 125 Hz. In their table they compare their results with the ones obtained in the literature for both untruncated and 125 Hz sampled datasets and truncated and 25 Hz sampled ones. See Ref. Table II in \cite{Mashhadi2015}.}& 1.72 & 1.33 & 0.90 & 1.28 & 0.93 & 1.41 & 0.61 & 0.88 & 0.59 & 3.78 & 0.85 & 0.71 & 1.25 & 0.87\\ \hline
{\tiny TROIKA - Zhang \textit{et al.} - 25 Hz \cite{Zhang2014}}\footnote{8s windows with 6s overlap. Sampling rate at 25 Hz. These values are provided in \cite{Khan2015}} & 3.05 & 3.49 & 1.49 & 2.03 & 1.46 & 2.35 & 1.76 & 1.42 & 1.28 & 5.73 & 1.79 & 3.02 & 2.41 & 1.28\footref{BHO}\\ \hline
{\tiny Khan \textit{et al.} - double channel - 25 Hz \cite{Khan2015}}\footref{8-6} & 1.64 & 0.81 & 0.57 & 1.44 & 0.77 & 1.06 & 0.63 & 0.47 & 0.52 & 2.94 & 1.05 & 0.91 & 1.07 & 0.69\footref{BHO}\\ \hline
{\tiny Khan \textit{et al.} - single channel - 25 Hz \cite{Khan2015}}\footref{8-6} & 2.55 & 3.45 & 0.73 & 1.19 & 0.51 & 1.09 & 0.52 & 0.43 & 0.36 & 3.33 & 0.89 & 0.98 & 1.34 & 1.12\footref{BHO}\\ \hline
{\tiny $\ours$\footnote{\label{L23_Troika}$\lambda = 0.023$} vs reference IHR}  &  3.00 & 3.06 & 2.98 & 2.26 & 2.47 & 2.71 & 2.49 & 3.55 & 2.92 & 5.70 & 1.80 & 2.76 & 2.97 & 0.97\\ \hline
{\tiny $\ours$-IF\footnote{\label{L21_Troika}$\lambda = 0.021$} vs reference AHR}  &  1.30 & 0.52 & 0.47 & 1.41 & 0.47 & 0.75 & 0.68 & 0.51 & 0.30 & 3.72 & 0.96 & 0.60 & 0.97 & 0.93\\ \hline
\hline\hline
 {\tiny AAEP (no unit)} & {\tiny Sbj 1} & {\tiny Sbj 2} & {\tiny Sbj 3} & {\tiny Sbj 4} & {\tiny Sbj 5} & {\tiny Sbj 6} & {\tiny Sbj 7} & {\tiny Sbj 8} & {\tiny Sbj 9} & {\tiny Sbj 10} & {\tiny Sbj 11} & {\tiny Sbj 12} & {\tiny mean} & {\tiny std}  \\ \hline\hline
{\tiny TROIKA - Zhang \textit{et al.} \cite{Zhang2014}}\footref{8-6}& 1.90 & 1.87 & 1.66 & 1.82 & 1.49 & 2.25 & 1.26 & 1.62 & 1.59 & 2.93 & 1.15 & 1.99 &  1.79 & 0.47\\ \hline
{\tiny SPECTRAP - Sun \textit{et al.} \cite{Sun2015}}\footref{8-6}\footref{Hz_U} & 1.04 & 2.33 & 0.66 & 1.31 & 0.74 & 1.14 & 1.36 & 0.55 & 0.52 & 2.27 & 0.65 & 1.02 & 1.13 & 0.61\footref{BHO}\\ \hline
{\tiny MICROST - Zhu \textit{et al.} \cite{Zhu_Tan:2015}}\footref{8-6}& 2.55 & 2.94 & 1.60 & 1.89 & 1.80 & 2.03 & 1.49 & 1.50 & 1.64 & 2.39 & 1.31 & 2.76 & 1.99\footref{WVM} & 0.54\\ \hline
{\tiny Frigo \textit{et al.} \cite{Frigo2015}}\footref{8-6}& 1.71 & 1.56 & 0.88 & 1.00 & 0.46 & 1.37 & 0.42 & 0.52 & 0.48 & 2.75 & 0.74 & 1.86 & 1.15 & 0.72\\ \hline
{\tiny D'souza \textit{et al.} \cite{SandeepJarChakraborti2015}}\footref{8-6} & 3.02 & 3.02 & 2.20 & 1.96 & 0.67 & 2.09 & 0.63 & 0.93 & 0.62 & 2.29 & 1.06 & 2.60 & 1.76 & 0.93\\ \hline
{\tiny Zhang \textit{et al.} \cite{ZhangLiuZhang2015a}}\footref{8-6}& 1.66 & 3.50 & 0.73 & 1.41 & 0.72 & 1.24 & 1.55 & 0.53 & 0.51 & 2.83 & 0.84 & 1.25 & 1.40 & 0.92\\ \hline
{\tiny Mashhadi \textit{et al.} \cite{Mashhadi2015}}\footref{Mashhadi2015}& 1.5 & 1.3 & 0.75 & 1.2 & 0.69 & 1.2 & 0.5 & 0.8 & 0.5 & 2.4 & 0.6 & 0.5 & 1.00 & 0.56\\ \hline
{\tiny $\ours$\footref{L23_Troika} vs reference IHR} & 2.85 & 3.01 & 2.50 & 1.93 & 1.89 & 2.33 & 2.01 & 3.22 & 2.60 & 3.84 & 1.23 & 2.17 & 2.47 & 0.70\\ \hline
{\tiny $\ours$-IF\footref{L21_Troika} vs reference AHR}  & 1.18 & 0.50 & 0.40 & 1.20 & 0.36 & 0.58 & 0.51 & 0.45 & 0.26 & 2.43 & 0.66 & 0.44 & 0.75 & 0.61\\ \hline
\end{tabular}
}
\caption{The average absolute error (AAE{, the same as the mean absolute error}) and average absolute error percentage (AAEP) of the heart rate estimation for the ICASSP 2015 signal processing cup database. The unit for the heart rate (HR) is beats per minute. We compare results and statistics of previously developed method with the ones of $\ours$ algorithm with and without a smoothing window of 8 seconds shifted of 2 seconds. Sbj: subject. std: standard deviation.}
\label{tab:Motion_HR_AAEP}
\end{minipage}
\end{table}

{
\subsection{Computational complexity}

The computational complexity of the proposed algorithm is linearly related to the length of the PPG signal. For the Capnobase benchmark database, using MATLAB R2011a installed on a 64-bit Windows 7 Professional computer equipped with a core i3-3227U CPU, 1.9 GHZ, and 8GB RAM, the average computational time for the whole algorithm to output the IHR and IRR is around 13 seconds which includes all steps in the algorithm. The only method in the literature for which we know the computational time is \cite{Zhu_Pimentel:2015} where the BCLA method is proposed. In that work, the authors state that ``the average time for fusing 900 estimates from six algorithms using BCLA was about 0.64 seconds''. At first glance, our algorithm is slow compared with the BCLA algorithm. However, note that the method in \cite{Zhu_Pimentel:2015} computes only respiratory rates. Furthermore, it is not clear if the reported time includes the computation of the 900 estimates which are necessary in the BCLA method. We note also that the signals that are provided at 300 Hz are re-sampled, before applying the BCLA method, at 4 Hz using linear interpolation.}

\section{Discussion}\label{Section:Discussion}

Given a PPG signal, recorded from a subject performing some activities, even intensive ones, the challenge is to extract at once different physiological dynamics, ideally instantaneous ones.
In this paper, we propose a new mathematically rigorous algorithm, named $\ours$, which is detailed in Section \ref{Section:Methods}, to achieve this goal. This method is able to extract the IHR and IRR simultaneously from a single PPG signal, {In order to clarify the meaning of IHR/IRR, we provide a discussion about the notion of ``instantaneous frequency'' from the measurement scale viewpoint.} 
%
and it proves to have better performance than any other method previously proposed and to properly extract IHR and IRR curves from a single channel PPG signal, even in the presence of strong motion artifacts.

Compared with the available methods in the literature, the main novelty of the proposed algorithm is threefold.
First, it allows us to simultaneously extract the IHR and IRR from the single-lead PPG signal, even when the cardiac and respiratory dynamical oscillations are both time-varying in frequencies and amplitudes, and are non-sinusoidally shaped. The key feature of the $\ours$ algorithm is a nonlinear mask design, which technically converts non-sinusoidal oscillations into sinusoidal ones. This enables us to simultaneously extract IHR and IRR curves.
Second, the proposed method is robust to nonstationary noise, and, even more importantly, to motion artifacts. Although not pursued in this work, based on the numerical results, we conjecture that the proposed method is also able to extract moving rhythms like, for example, the gait.
Third, the algorithm is local in nature, which allows us to better extract the finer dynamical structure. The non-adaptive truncation-and-stitch policy commonly employed in the field is no longer needed.

{There are at least two sources leading to the error in the semi-real PPG simulation database that are worth mentioning. 
The limitation of de-shape type method plays a fundamental role in the $\ours$ algorithm. Based on the existing theory shown in \cite{lin2016waveshape}, when the variation of the IHR is ``dramatic'', the accuracy of the IHR estimation will be downgraded. However, it is well known that the IHR could have sudden changes due to the fractal Brownian motion behavior of the RRI time series. A consequence of this limitation is a possibly oversmoothed estimated IHR, which leads to the first source of the error (the orange arrows in Figure \ref{fig:Simulation}). 
Furthermore, the standardized PPG cycle set is constructed in a way that we could only preserve the continuity property of the PPG signal in the simulated PPG signal, but no guarantee about the smoothness. Particularly, the phase of the simulated PPG signal might not be smooth, which leads to the second source of error. 
Note that although the second source of error could be improved by designing a better simulation model, we cannot improve the first source of error since it is rooted in the current theoretical limitation. As a result, we only obtain an instantaneous information up to the scale of around $20$ msec, which could be roughly related to the ECG signal sampled at $50$ Hz. While developing a new theoretical work and HRV analysis are out of the scope of this paper, we leave a further theoretical and algorithmic study to the future.

The accuracy of $\ours$ in the benchmark databases also deserves a discussion. In this study, we take the IHR determined from the RRI as the gold standard, and take the IHR estimated from the PPG by $\ours$ as an estimate. 
%
While intuitively the estimate seem reasonable since the oscillations in the PPG signal are caused by the cardiac activities,
however, the estimation cannot be that accurate due to the pulse transit time (PTT), which is a fundamental physiological relationship between the ECG and PPG signals. 
In short, there is a few hundred milliseconds latency between the ventricular response and the PPG pulse observed right after that ventricular response in the ECG waveform. This latency comes from the fact that the pumped blood from the heart takes time to travel through the arteries and capillaries in order to arrive at the PPG measurement location. 
Even worse, it has been widely known that the PTT depends on the hemodynamics, and it varies from time to time and provides another set of information \cite{Kim2013}. 
As a result, due to the PTT, the IHR estimated from the PPG might deviate from the gold standard IHR determined from the RRI. In the real scenario, since there is no way to obtain the PTT when we only have a single channel PPG, an error is inevitable. How to obtain a more accurate IHR estimate is an important topic that will be further explored in the future work.
}

The noise-robustness property of the proposed $\ours$ further broadens its applicability. Recently, due to the advancements of light source generation, more and more effort has been put into the non-contact PPG analysis. As the terminology suggests, unlike the PPG signal, the non-contact PPG is collected from a subject without any direct contact; for example, the PhysioCam \cite{Davlia_Lewis_Porges:2016}, or other methods that collect the video from a subject \cite{Davlia_Lewis_Porges:2016,McDuff2016}. For those techniques, the collected video is post-processed by the researcher to generate the PPG signal. In general, these kinds of signals are noisier, and the information content is not fully understood, except the cardiac dynamics. {Clearly, for this kind of signal, the landmark based approach to estimate the ``beat-to-beat interval'' is even more difficult}. The proposed method has the potential to be applied to this kind of signal. Specifically, the synchrosqueezing transform has been applied to study {the non-contact PPG signal from the PhysioCam technology} \cite{Wu_Lewis_Davila_Daubechies_Porges:2015} and a preliminary analysis result by the de-shape synchrosqueezing transform has been shown in \cite[Section 4.3.3]{lin2016waveshape}.

The discussion cannot be complete without mentioning the directions of possible further improvement of the proposed algorithm.
First of all, while $\ours$ performs well overall, in general the performance of curve extraction step can be further improved. How to design an efficient automatic curve extraction algorithm is an open problem, and several efforts have been invested so far in this direction, for example \cite{6265417}. In this work we adopt a basic method, based on a minimization approach, to get the result. Having a more efficient and automatic algorithm, would allow to obtain even better results.

Second, in the $\ours$ algorithm we sharpen the {nonlinearly masked} spectrogram by taking into account the phase information computed by the STFT. However, there are several other possibilities to extract more detailed information from the phase function like, for example, the second-order synchrosqueezing transform \cite{Oberlin2015}. While we do not exhaustively explore the possibility in this work, it is promising to combine it into the $\ours$ algorithm. 

Third, while the databases we analyze are publicly available and close to several real scenarios, the database size is not large enough from a statistical point of view. Furthermore, not all possible physiological information, physical activity details and environmental conditions are included. A well-designed prospective study with different setups is needed to further evaluate the proposed algorithm.

Fourth, we do not extract all of the available information contained in the PPG signal. In addition to the RIIV, respiratory dynamics are hidden in different forms inside a PPG signal, like the pulse-wave transit time, the PPG pulse width and the respiratory sinus arrhythmia. This means that we could extract several different respiratory signals from a single-lead PPG signal. Furthermore, if we are allowed to also use an ECG channel, then the pulse-wave transit time and the ECG derived respiratory signal could be extracted and used to further improve the results. 

Finally, we want to mention that the proposed algorithm has unexplored potential. In this work, we study the PPG signal in order to estimate IHR and IRR curves, so that the $\ours$ algorithm stops once the curves have been extracted from the de-shaped spectrogram. However, for some other applications, we could continue analyzing the PPG signal. It would be possible to perform the wave-shape reconstruction, count the oscillatory components, decompose each oscillatory component, etc. Furthermore, we could even apply the proposed method to other signals. For example, in the fetal ECG signal analysis, decomposing the maternal ECG signal from the fetal ECG signal is a critical step \cite{Su_Wu:2016b}.

\section*{Acknowledgements}

Antonio Cicone research is supported by Istituto Nazionale di Alta Matematica (INdAM) ``INdAM Fellowships in Mathematics and/or Applications cofunded by Marie Curie Actions'', PCOFUND-GA-2009-245492 INdAM-COFUND Marie Sklodowska Curie Integration Grants. Hau-tieng Wu's research is partially supported by Sloan Research Fellow FR-2015-65363.
The authors want to thank Professor Li Su for sharing the initial implementation of the de-shape SST algorithm, {and Dr. Yu-Ting Lin for sharing the PPG signal recorded from subjects under general anesthesia for the simulation.}

\bibliographystyle{amsplain}
\bibliography{PPGCepstrum_v3}

\appendix
\renewcommand{\thefigure}{SI.\arabic{figure}}
\setcounter{figure}{0}
\renewcommand{\thetable}{SI.\arabic{table}}
\setcounter{table}{0}
\renewcommand{\theequation}{S.\arabic{equation}}
\setcounter{equation}{0}
\renewcommand{\thesection}{SI.\arabic{section}}
\setcounter{section}{0}

{\LARGE \center \textbf{Appendix}\\}

\bigskip

\section{More details about the results}\label{supp:MoreResults}

\subsection{Capnobase benchmark database -- PPG signal with respiration}\label{supp:Capnobase}

\begin{figure}
\centering
\includegraphics[width=0.9\columnwidth]{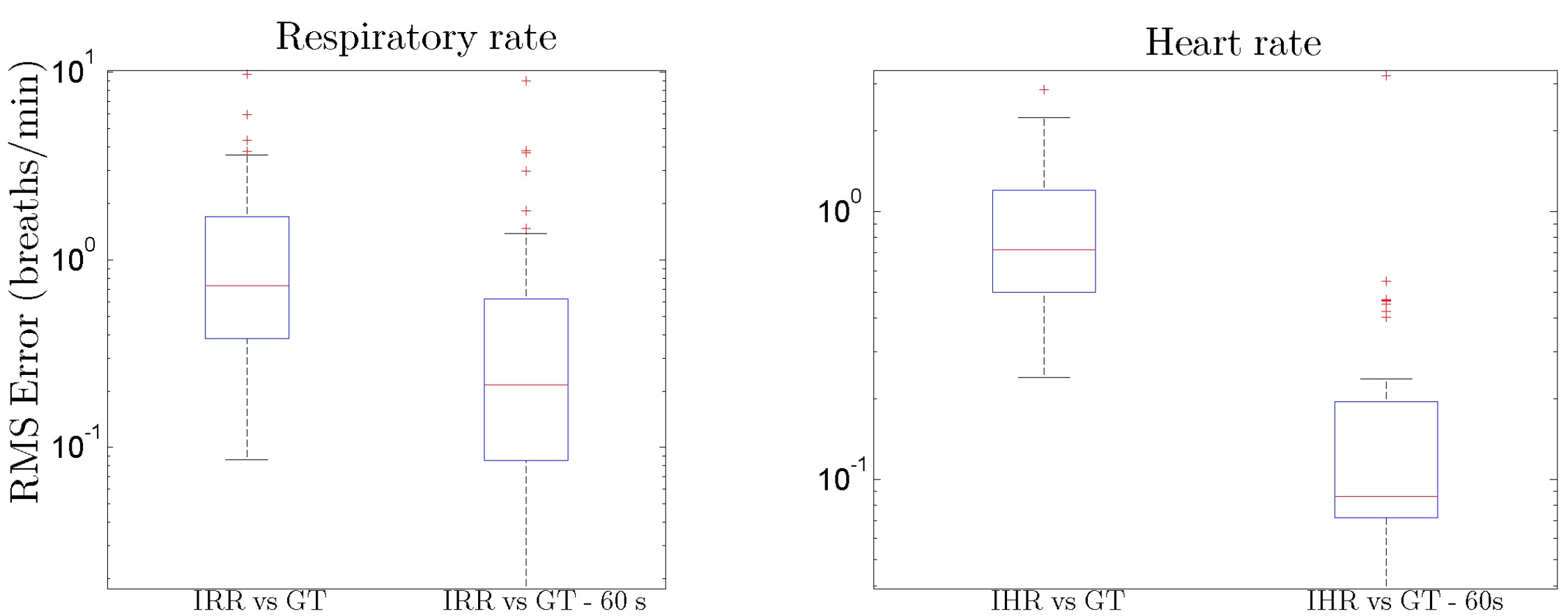}
\caption{The boxplots of the root mean square error (RMS) of the instantaneous respiratory and the instantaneous heart rate estimation are shown on the left and right panel respectively. They have been generated using the Matlab function \texttt{boxplot}. The boxplots include the RMS of the instantaneous values or the average value over a 60 seconds window versus the given reference or ground truth (GT). The first, second, and third quartiles are displayed as bottom, middle, and top horizontal line of the boxes. Whiskers represents the most extreme values within three times the interquartile range from the quartile. Crosses represents outliers.}
\label{fig:Capnobase_boxplot_givenRef}
\end{figure}

As is shown in Table 1 in the main article, the performance of the $\ours$ method is better than other methods proposed in the literature.
To better understand such performance, we report also the boxplots in Figure \ref{fig:Capnobase_boxplot_givenRef}.

The Capnobase benchmark database provides labels regarding intervals containing potential artifacts in the PPG, ECG and capnometry signals. Other methods proposed in the literature, to the best of our knowledge, always exploit this information to remove these intervals in computing the statistics.
However, it is important to note that to address the need for automatic annotation of respiratory and heart rates in a real scenario, the performance of $\ours$ is obtained disregarding all the information about the intervals containing potential artifacts: we do use all the 42 datasets in the Capnobase benchmark database and all their intervals, even the ones known to contain artifacts.

\begin{figure}
\centering
\includegraphics[width=0.9\columnwidth]{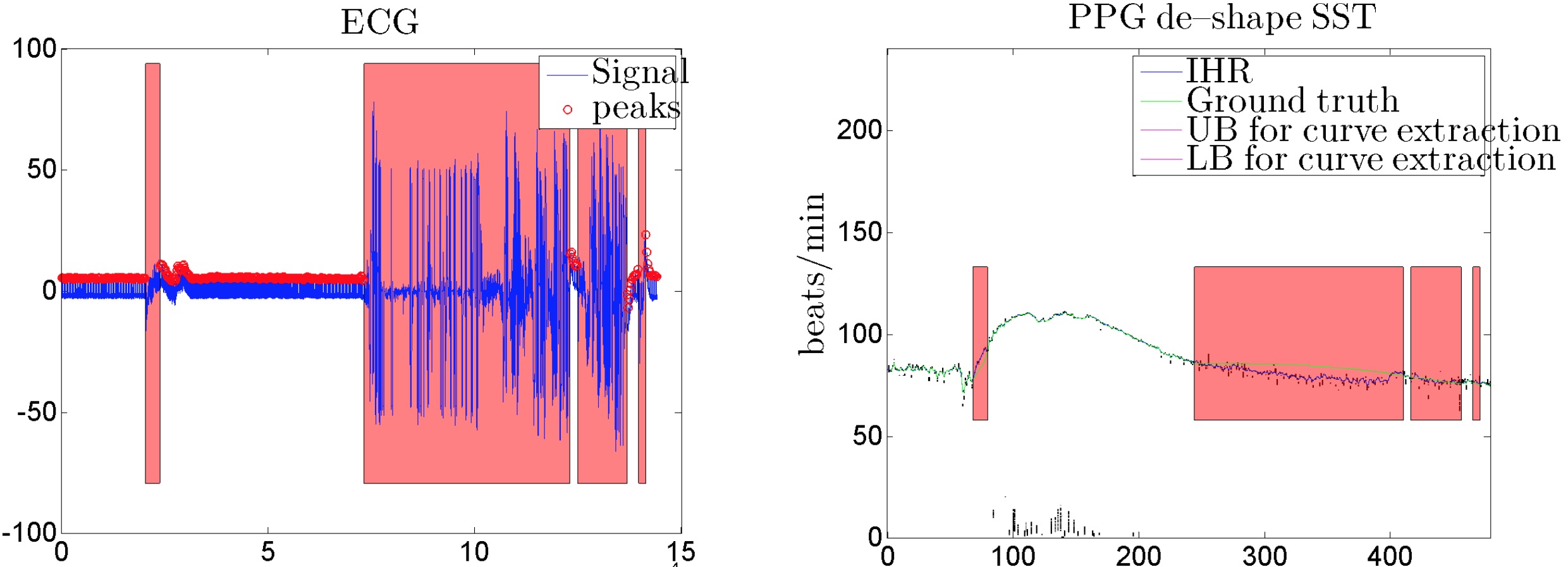}
\caption{Dataset \texttt{0329\textunderscore 8min}. The electrocardiogram (ECG) signal is shown on the left panel. Whereas, on the right panel, it is shown the de-shaped spectrogram of the PPG signal that is superimposed with the reference averaged heart rate and the extracted curve corresponding to the instantaneous heart rate. The red boxes highlight the intervals reported in the database as containing artifacts. The reference curve is plotted in solid red. LB: lower boundary; UB: upper boundary.}\label{fig:Capno_Artifacts2}
\end{figure}

On the other hand, the reference curves provided in the database may not be reliable in the intervals containing such artifacts. For instance in Figure \ref{fig:Capno_Artifacts2} we report for the dataset \texttt{0329\textunderscore 8min} the ECG signal, the de-shaped spectrogram as well as the reference and extracted curves corresponding to IHR. Here, it is evident that the reference AHR curve, which is marked in solid red, has been produced using some kind of interpolation for the interval 244 - 410 seconds that has been reported as containing artifacts.

In order to have a better comparison with the performance of other methods, we recompute the root mean square (RMS) error and mean absolute error (MAE) of the proposed $\ours$ and $\ours$-60s methods, when points inside the intervals containing artifacts are not taken into account. The results are shown in Table \ref{tab:Capnobase_NoArtifacts} as ``artifacts removed''.
If we compare these statistics with the ones obtained without removing the artifact intervals, we observe that for the HR, we have better performance in the instantaneous case, while the performance does not change in the average case.
Regarding the respiratory rate, instead, there are no changes at all in the performance both for the instantaneous and average cases.

\begin{figure*}
\centering
\includegraphics[width=0.9\columnwidth]{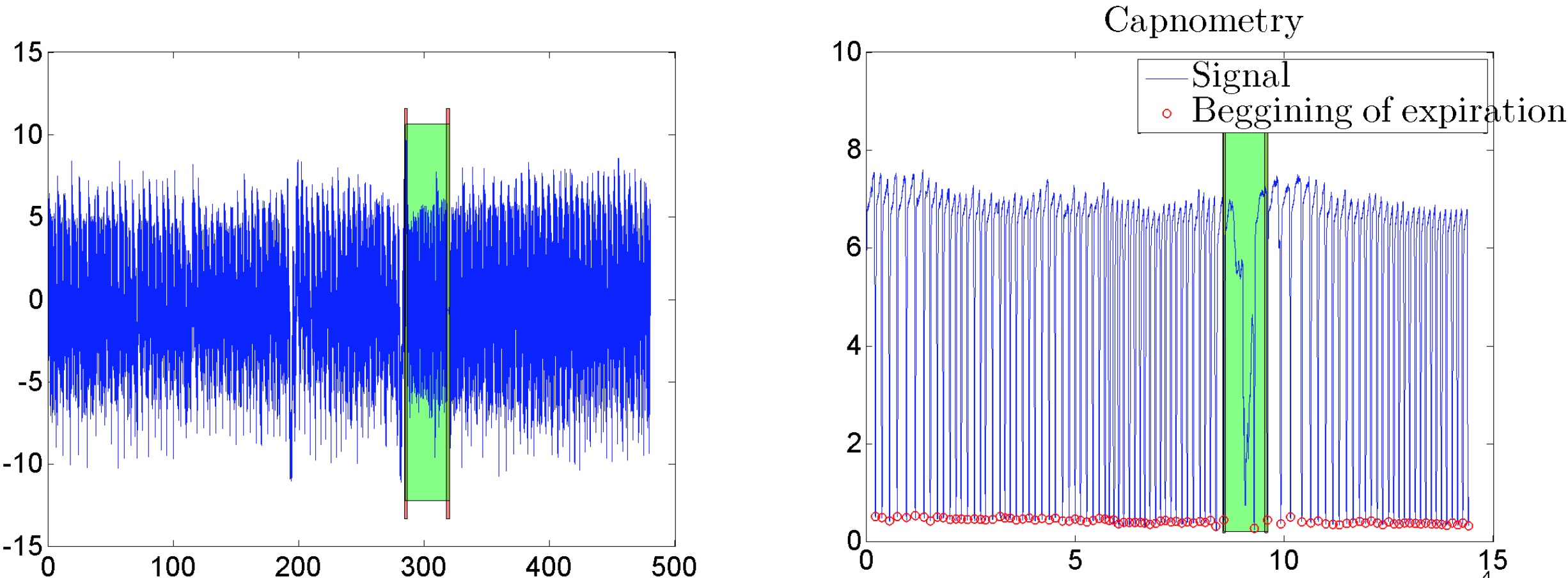}
\caption{The PPG signal of the dataset \texttt{0032\textunderscore 8min} is shown on the left, and the simultaneously recorded capnometry signal is shown on the right. Red boxes highlight intervals reported in the database as containing artifacts. The green box highlights the newly identified interval in the capnometry containing artifacts.}\label{fig:Capno_Artifacts}
\end{figure*}

To better understand this phenomenon, we take a close look at the 42 datasets and their intervals containing artifacts. We discover that there are four datasets for which the beginning of expiration values are missing in one or more intervals which have not been labeled as containing artifacts. An example is shown in Figure \ref{fig:Capno_Artifacts}, where the PPG signal and the capnometry for \texttt{0032\textunderscore 8min} is reported. On the other hand, the artificial intervals of the ECG signal are all well labeled.
We provide complete names of these datasets and the corresponding intervals, denoted using sample point positions of their boundaries:

\texttt{0031\textunderscore 8min} $[1, \, 38665]$, $[39644, \, 67316]$, $[71295, \, 76682]$, $[83126, \, 144001]$,

\texttt{0032\textunderscore 8min} $[85445, \, 95970]$,

\texttt{0328\textunderscore 8min} $[116923, \, 124135]$,

\texttt{0370\textunderscore 8min} $[15271, \, 24817]$, $[46804, \, 78243]$.

If we also take into account these new intervals in removing artifacts for the computation of the $\ours$ method performance, we obtain the statistics shown in Table \ref{tab:Capnobase_NoArtifacts} as ``artifacts removed, updated''. As expected, the performance for the HR estimation does not change. However, for the RR we have a slight improvement. This extra performance evaluation indicates the importance of taking the signal quality index (SQI) \cite{Gambarotta2016,Fischer2016} into account. For any practical application, it is clear that, in general, we do not have any ground truth and experts are not available to determine which interval is of low quality. We could than apply the SQI designed for the PPG signal, the ECG signal, and so on, to determine which interval can be trusted. Since the SQI is not the focus of this paper, we plan to report the research result by taking the SQI into account in the $\ours$ application to a signal in a future work.

\begin{table*}
\footnotesize
\begin{minipage}{\textwidth}
\begin{tabular}{|c|c||c|c|c|c|c||c|c|c|c|c|}\hline
\multirow{2}{*}{Given Reference}  & \multirow{2}{*}{HR (beats/minute)} & \multicolumn{5}{c||}{RMS} & \multicolumn{5}{c|}{MAE} \\ \cline{3-12}
  & & mean & std & $Q_1$ & median & $Q_3$  & mean & std & $Q_1$ & median & $Q_3$ \\ \hline
\hline
\multirow{3}{*}{$\ours$} & whole database & 0.93 & 0.57 & 0.50 & 0.72 & 1.20 & 0.61 & 0.35 & 0.38 & 0.52 & 0.84\\ \cline{2-12}
& artifacts removed & 0.79 & 0.41 & 0.50 & 0.67 & 1.06 & 0.55 & 0.27 & 0.37 & 0.48 & 0.76\\ \cline{2-12}
& artifacts removed, updated & 0.79 & 0.43 & 0.46 & 0.67 & 1.06 & 0.56 & 0.28 & 0.36 & 0.48 & 0.77\\ \hline
\multirow{3}{*}{$\ours$-60s} & whole database & 0.23 & 0.49 & 0.07 & 0.09 & 0.19 & 0.15 & 0.29 & 0.05 & 0.07 & 0.12\\ \cline{2-12}
& artifacts removed & 0.24 & 0.54 & 0.07 & 0.09 & 0.19 & 0.16 & 0.32 & 0.05 & 0.07 & 0.12\\ \cline{2-12}
& artifacts removed, updated & 0.24 & 0.54 & 0.07 & 0.09 & 0.19 & 0.16 & 0.32 & 0.05 & 0.07 & 0.12\\ \hline\hline\hline
\multirow{2}{*}{Given Reference} & \multirow{2}{*}{RR (breaths/minute)} & \multicolumn{5}{c||}{RMS} & \multicolumn{5}{c|}{MAE} \\ \cline{3-12}
 & & mean & std & $Q_1$ & median & $Q_3$  & mean & std & $Q_1$ & median & $Q_3$ \\ \hline
\hline
\multirow{3}{*}{$\ours$} & whole database  & 1.39 & 1.87 & 0.38 & 0.73 & 1.70 & 0.94 & 1.37 & 0.22 & 0.50 & 0.83\\ \cline{2-12}
& artifacts removed & 1.39 & 1.88 & 0.38 & 0.73 & 1.70 & 0.94 & 1.39 & 0.22 & 0.50 & 0.83\\ \cline{2-12}
& artifacts removed, updated & 1.37 & 1.90 & 0.38 & 0.73 & 1.70 & 0.91 & 1.33 & 0.22 & 0.50 & 0.83\\ \hline
\multirow{3}{*}{$\ours$-60s} & whole database  & 0.78 & 1.60 & 0.09 & 0.22 & 0.62 & 0.53 & 1.16 & 0.07 & 0.15 & 0.43\\ \cline{2-12}
& artifacts removed & 0.78 & 1.60 & 0.09 & 0.22 & 0.62 & 0.53 & 1.16 & 0.07 & 0.15 & 0.43\\ \cline{2-12}
& artifacts removed, updated & 0.81 & 1.74 & 0.09 & 0.22 & 0.62 & 0.56 & 1.32 & 0.07 & 0.15 & 0.43\\ \hline
\end{tabular}
\caption{The root mean square error (RMS) and mean absolute error (MAE) of the respiratory rate (RR) and heart rate (HR) estimation obtained by the $\ours$ method for the Capnobase benchmark database in the following three scenarios. The estimated RR and HR are compared with the reference curves provided in the database.
In the lines labeled ``whole database'' we provide performance of the methods based on the whole database without removing any interval. The results are the same as those shown in Table \ref{tab:Capno_stats}.
Second, in the lines labeled ``artifacts removed'', we provide results when skipping labeled intervals containing artifacts provided in the database.
Third, in the lines labeled ``artifacts removed, updated'' we provide performance when we further skip intervals for which no beginning of expiration information are provided, in addition to skipping intervals labeled as containing artifacts.
The unit of the HR is beats per minute, whereas the unit of RR is breaths per minute. Std: standard deviation. $Q_1$ and  $Q_3$: first and third quartile.}
\label{tab:Capnobase_NoArtifacts}
\end{minipage}
\end{table*}

Another important aspect to point out is that the reference curves provided are intended to be used for estimating AHR and ARR over a time window, whereas the proposed $\ours$ method allows the computation of the \textit{instantaneous} respiratory and heart rates. In order to ensure that such curves are also reliable for evaluating the instantaneous performance of the proposed $\ours$ method, we want to compute new instantaneous reference curves. We compute them using expiration beginnings and R-peaks position, which are also provided in these datasets. In particular, in each time instant in the middle of two consecutive expirations in the capnogram signal (respectively two consecutive R-peaks in the ECG signal), we compute the reciprocal of the length of the time interval between these two expirations (respectively R-peaks). We then derive the IRR (respectively IHR) as the cubic spline interpolation of such values.
In Table \ref{tab:Capnobase_NoArtifactsInstantaneous}, we report the performance of the $\ours$ method computed using the newly computed IRR and IHR as reference curves.

\begin{table*}
\footnotesize
\begin{minipage}{\textwidth}
\begin{tabular}{|c|c||c|c|c|c|c||c|c|c|c|c|}\hline
\multirow{2}{*}{New Reference} & \multirow{2}{*}{HR (beats/minute)} & \multicolumn{5}{c||}{RMS} & \multicolumn{5}{c|}{MAE} \\ \cline{3-12}
 & & mean & std & $Q_1$ & median & $Q_3$  & mean & std & $Q_1$ & median & $Q_3$ \\ \hline
\hline
\multirow{3}{*}{$\ours$} & whole database & 2.76 & 5.59 & 0.76 & 0.97 & 1.56 & 1.38 & 3.12 & 0.54 & 0.69 & 1.10\\ \cline{2-12}
& artifacts removed & 1.01 & 0.56 & 0.67 & 0.85 & 1.28 & 0.72 & 0.34 & 0.52 & 0.64 & 0.92\\ \cline{2-12}
& artifacts removed, updated & 1.02 & 0.58 & 0.67 & 0.85 & 1.28 & 0.72 & 0.35 & 0.52 & 0.64 & 0.92\\ \hline
\multirow{3}{*}{$\ours$-60s} & whole database & 1.16 & 4.42 & 0.05 & 0.07 & 0.25 & 0.71 & 2.87 & 0.05 & 0.06 & 0.17\\ \cline{2-12}
& artifacts removed & 0.82 & 2.36 & 0.05 & 0.07 & 0.25 & 0.49 & 1.50 & 0.05 & 0.06 & 0.17\\ \cline{2-12}
& artifacts removed, updated & 0.82 & 2.36 & 0.05 & 0.07 & 0.28 & 0.50 & 1.50 & 0.05 & 0.06 & 0.19\\ \hline\hline\hline
\multirow{2}{*}{New Reference} & \multirow{2}{*}{RR (breaths/minute)} & \multicolumn{5}{c||}{RMS} & \multicolumn{5}{c|}{MAE} \\ \cline{3-12}
 && mean & std & $Q_1$ & median & $Q_3$  & mean & std & $Q_1$ & median & $Q_3$ \\ \hline\hline
\multirow{3}{*}{$\ours$}  & whole database & 2.42 & 9.14 & 0.28 & 0.54 & 1.16 & 1.17 & 3.94 & 0.18 & 0.40 & 0.77\\ \cline{2-12}
& artifacts removed & 2.45 & 9.47 & 0.26 & 0.54 & 1.15 & 1.20 & 4.18 & 0.18 & 0.39 & 0.77\\ \cline{2-12}
& artifacts removed, updated & 1.04 & 1.73 & 0.26 & 0.52 & 0.91 & 0.65 & 1.12 & 0.18 & 0.37 & 0.57\\ \hline
\multirow{3}{*}{$\ours$-60s} & whole database & 2.87 & 15.59 & 0.04 & 0.08 & 0.49 & 1.24 & 6.21 & 0.04 & 0.06 & 0.25\\ \cline{2-12}
& artifacts removed & 2.87 & 15.59 & 0.05 & 0.08 & 0.49 & 1.24 & 6.21 & 0.04 & 0.06 & 0.25\\ \cline{2-12}
& artifacts removed, updated & 0.71 & 1.84 & 0.05 & 0.08 & 0.49 & 0.41 & 1.02 & 0.04 & 0.06 & 0.25\\ \hline
\end{tabular}
\caption{The root mean square error (RMS) and mean absolute error (MAE) of the instantaneous respiratory rate (IRR) and instantaneous heart rate (IHR) estimation obtained by the $\ours$ method for the Capnobase benchmark database in the following three scenarios. The estimated IRR and IHR are compared with a new reference IRR (respectively IHR) determined directly from the beginning of expirations in the capnogram signal (respectively the R-peaks of the electrocardiogram signal).
Second, in the lines labeled ``whole database'' we provide results based on the whole database without removing any interval.
Whereas in the lines labeled ``artifacts removed'', we provide results when skipping labeled intervals containing artifacts provided in the database.
Third, finally in the lines labeled ``artifacts removed, updated'', we provide performance when we further skip intervals for which no beginning of expiration information are provided, in addition to skipping intervals labeled as containing artifacts.
The unit of the HR is beats per minute, whereas the unit of RR is breaths per minute. Std: standard deviation. $Q_1$ and  $Q_3$: first and third quartile.
}
\label{tab:Capnobase_NoArtifactsInstantaneous}
\end{minipage}
\end{table*}

Regarding the performance of the $\ours$  method when we use the newly generated reference curves, it has to be taken into account that in some datasets the R-peaks and expiration beginnings positions are not provided for the entire signal. Nevertheless, we compute the new reference curves without making any assumption on the datasets and using the provided R-peaks and expiration beginnings as if they were reliable in all intervals.
Clearly, for datasets containing artifacts, the newly generated reference curves tend to have wrong values which, in turn, impact negatively on the computed values of the proposed method's performance. In particular, the datasets containing artifacts have high RMS and MAE when we use the new reference curves. This increases the statistical values used to measure the performance of $\ours$. The proposed method would have better performance in terms of RMS and MAE, if more reliable information on the R-peaks and expiration beginnings positions were available. This becomes more evident when we remove the intervals containing artifacts labeled in the database, ``artifacts removed'' case in Table \ref{tab:Capnobase_NoArtifactsInstantaneous}, as well as the newly identified intervals with artifact in the capnometry signal mentioned before, ``artifacts removed, updated'' case in the same Table.

In summary, from Tables \ref{tab:Capnobase_NoArtifacts} and \ref{tab:Capnobase_NoArtifactsInstantaneous}, we see that the best performance for the RR are obtained using the new reference curves and if we remove all the artifacts, including the ones not originally included in the database, whereas the performance for the HR are better when we compare the estimated curves with the given reference curves and after removing the artifacts.

\subsection{ICASSP 2015 signal processing cup -- PPG signal with motion}\label{supp:ICASSP}

For each dataset, the ground-truth of heart rate, saved as variable 'BPM0', is provided in the database and can be calculated from the simultaneously recorded ECG signal. Since our purpose is evaluating the IHR, it is important to detail how the reference AHR provided in the database is calculated \cite{Zhang2014}. In particular, the reference AHR is computed using a time window of $D = 8$ seconds in the following way. First, the number of cardiac cycles, denoted as $H$, is counted over time windows of length $D$ (in seconds), the HR in each window is computed as $60\times H/D$, which has unit beats per minute (BPM). Two successive time windows overlap by 6 seconds. Therefore, the first value in 'BPM0' gives the calculated heart rate ground-truth in the first 8 seconds, while the second value in 'BPM0' gives the calculated heart rate ground-truth from the 3rd second to the 10th second. The reference HR could thus be understood as the smoothed IHR by a zero-one window eight seconds in length, and this smoothing effect explains why we need to smooth out our estimated curves using Iterative Filtering (IF) \cite{cicone2016adaptive,cicone2016hyperspectral,cicone2017MIF}, as explained previously, in order to obtain a good fitting result.

There is another important fact regarding the signal quality issue. Zhang in \cite{Zhang2015} proposed to truncate some of the datasets contained in the ICASSP Signal Processing Cup database in order to remove some motion artifacts. The excluded segments in that work
were: the first 12 seconds of Set 2, the first 8 seconds of Set 3, the first 2 seconds of Set 4, the first 2 seconds of Set 8, the first 6 seconds of Set 10 and the first 2 seconds of Set 11.
Furthermore, the same author proposed to downsample the PPG signals from 125 Hz to 25 Hz. We compare the performance of the $\ours$ method when applied to truncated and downsampled signals in Table \ref{tab:Motion_HR_AAEP_Truncated}. From these results, the downsampling technique does not seem to help the $\ours$ algorithm. In Table \ref{tab:Motion_HR_AAEP_Truncated}, we also show the AHR determined by the $\ours$-8s algorithm. We could see that the performance of the $\ours$-8s algorithm is slightly worse than the $\ours$-IF algorithm.

\begin{table*}
\begin{minipage}{\textwidth}
{\tiny
\begin{tabular}{|c||c|c|c|c|c|c|c|c|c|c|c|c||c|c|}\hline
  AAE & {\tiny Sbj 1} & {\tiny Sbj 2} & {\tiny Sbj 3} & {\tiny Sbj 4} & {\tiny Sbj 5} & {\tiny Sbj 6} & {\tiny Sbj 7} & {\tiny Sbj 8} & {\tiny Sbj 9} & {\tiny Sbj 10} & {\tiny Sbj 11} & {\tiny Sbj 12} & mean & std  \\ \hline
\hline
{\tiny TROIKA - Zhang \textit{et al.} \cite{Zhang2014}}\footnote{\label{8-6-25Hz}8s windows with 6s overlap}\footnote{\label{CiteZhang2015}Values reported in \cite{Zhang2015}} & 2.87 & 2.75 & 1.91 & 2.25 & 1.69 & 3.16 & 1.72 & 1.83 & 1.58 & 4.00 & 1.96 & 3.33 & 2.42 & 0.78\footnote{\label{BHO2}This value has been recomputed using Matlab \textit{std} function and does not match the value reported in the original paper.}\\ \hline
{\tiny JOSS - Zhang \cite{Zhang2015}\footref{8-6-25Hz}} & 1.33 & 1.75 & 1.47 & 1.48 & 0.69 & 1.32 & 0.71 & 0.56 & 0.49 & 3.81 & 0.78 & 1.04 & 1.29 & 0.90\footref{BHO2}\\ \hline
{\tiny Khan \textit{et al.} - double channel \cite{Khan2015}\footref{8-6-25Hz}} & 1.70 & 0.84 & 0.56 & 1.15 & 0.77 & 1.06 & 0.63 & 0.53 & 0.52 & 2.56 & 1.05 & 0.91 & 1.02 & 0.59\footref{BHO2}\\ \hline
{\tiny Khan \textit{et al.} - single channel \cite{Khan2015}\footref{8-6-25Hz}} & 1.77 & 1.94 & 0.73 & 1.19 & 0.51 & 1.09 & 0.52 & 0.43 & 0.36 & 3.43 & 0.89 & 0.98 & 1.15 & 0.88\footref{BHO2}\\ \hline
{\tiny Khan \textit{et al.} - double channel - 125 Hz \cite{Khan2015}\footref{8-6-25Hz}} & 1.83 & 0.85 & 0.63 & 1.21 & 0.65 & 1.03 & 0.70 & 0.50 & 0.47 & 2.83 & 1.14 & 0.90 & 1.06 & 0.67\footref{BHO2}\\ \hline
{\tiny Temko - double channel \cite{Temko2015}}\footnote{8s windows with 6s overlap. It is not clear if in \cite{Temko2015} the datasets are truncated or not. In fact, in that paper values are compared with both the TROIKA results for the untruncated database \cite{Zhang2014} and the JOint Sparse Spectrum (JOSS) ones related to the truncated one \cite{Zhang2015}. } & 1.23 & 1.26 & 0.72 & 0.98 & 0.75 & 0.91 & 0.67 & 0.91 & 0.54 & 2.61 & 0.94 & 0.98 & 1.04 & 0.54\footref{BHO2}\\ \hline
{\tiny $\ours$ - 125Hz\footnote{\label{L25}$\lambda = 0.025$} vs ref IHR} &  3.00 & 6.24 & 4.85 & 3.07 & 2.47 & 2.69 & 2.49 & 3.55 & 2.93 & 5.94 & 2.31 & 2.75 & 3.53 & 1.37\\ \hline
{\tiny $\ours$-IF - 125Hz\footnote{\label{L23}$\lambda = 0.023$} vs ref AHR} &  1.29 & 0.55 & 0.50 & 0.67 & 0.47 & 0.75 & 0.68 & 0.52 & 0.30 & 3.32 & 0.52 & 0.60 & 0.85 & 0.81\\ \hline
{\tiny $\ours$-8s - 125Hz\footref{L23}\footref{8-6-25Hz} vs ref AHR} & 1.31 & 0.90 & 0.77 & 0.93 & 0.75 & 0.86 & 0.75 & 0.77 & 0.67 & 3.28 & 0.70 & 0.80 & 1.04 & 0.73\\ \hline
{\tiny $\ours$ - 25Hz\footnote{\label{L38}$\lambda = 0.038$} vs ref IHR}  & 2.80 & 5.42 & 4.84 & 3.37 & 2.58 & 2.71 & 2.61 & 3.52 & 2.91 & 6.37 & 2.31 & 2.80 & 3.52 & 1.31\\ \hline
{\tiny $\ours$-IF - 25Hz\footnote{\label{L48}$\lambda = 0.048$} vs ref AHR} & 0.87 & 0.57 & 0.51 & 1.15 & 0.62 & 0.73 & 0.76 & 0.52 & 0.29 & 3.97 & 0.50 & 0.83 & 0.94 & 0.98\\ \hline
{\tiny $\ours$-8s - 25Hz\footref{L48}\footref{8-6-25Hz} vs ref AHR} & 1.02 & 0.94 & 0.83 & 1.42 & 0.90 & 0.85 & 0.84 & 0.80 & 0.65 & 3.95 & 0.70 & 1.06 & 1.16 & 0.90\\ \hline
\hline\hline
  AAEP & {\tiny Sbj 1} & {\tiny Sbj 2} & {\tiny Sbj 3} & {\tiny Sbj 4} & {\tiny Sbj 5} & {\tiny Sbj 6} & {\tiny Sbj 7} & {\tiny Sbj 8} & {\tiny Sbj 9} & {\tiny Sbj 10} & {\tiny Sbj 11} & {\tiny Sbj 12} & mean & std  \\ \hline
\hline
{\tiny TROIKA - Zhang \textit{et al.} \cite{Zhang2014}\footref{8-6-25Hz}\footref{CiteZhang2015}} & 2.18 & 2.37 & 1.50 & 2.00 & 1.22 & 2.51 & 1.27 & 1.47 & 1.28 & 2.49 & 1.29 & 2.30 & 1.82 & 0.53\footref{BHO2}\\ \hline
{\tiny JOSS - Zhang \cite{Zhang2015}\footref{8-6-25Hz}} & 1.19 & 1.66 & 1.27 & 1.41 & 0.51 & 1.09 & 0.54 & 0.47 & 0.41 & 2.43 & 0.51 & 0.81 & 1.02 & 0.61\footref{BHO2}\\ \hline
{\tiny $\ours$ - 125Hz\footref{L25} vs ref IHR} & 2.85 & 6.01 & 4.06 & 2.66 & 1.89 & 2.32 & 2.01 & 3.23 & 2.61 & 3.96 & 1.60 & 2.16 & 2.95 & 1.23\\ \hline
{\tiny $\ours$-IF - 125Hz\footref{L23} vs ref AHR} & 1.17 & 0.54 & 0.43 & 0.64 & 0.36 & 0.58 & 0.51 & 0.46 & 0.26 & 2.11 & 0.35 & 0.44 & 0.65 & 0.51\\ \hline
{\tiny $\ours$-8s - 125Hz\footref{L23}\footref{8-6-25Hz} vs ref AHR} & 1.16 & 0.86 & 0.63 & 0.83 & 0.56 & 0.67 & 0.56 & 0.67 & 0.58 & 2.08 & 0.47 & 0.58 & 0.80 & 0.44\\ \hline
{\tiny $\ours$ - 25Hz\footref{L38} vs ref IHR} & 2.68 & 4.89 & 4.05 & 3.07 & 1.96 & 2.34 & 2.09 & 3.19 & 2.60 & 4.38 & 1.60 & 2.19 & 2.92 & 1.04\\ \hline
{\tiny $\ours$-IF - 25Hz\footref{L48} vs ref AHR} & 0.83 & 0.56 & 0.44 & 1.24 & 0.50 & 0.56 & 0.56 & 0.46 & 0.25 & 2.74 & 0.33 & 0.64 & 0.76 & 0.67\\ \hline
{\tiny $\ours$-8s - 25Hz\footref{L48}\footref{8-6-25Hz} vs ref AHR} & 0.93 & 0.90 & 0.68 & 1.42 & 0.70 & 0.67 & 0.62 & 0.70 & 0.57 & 2.71 & 0.47 & 0.80 & 0.93 & 0.61\\ \hline
\end{tabular}
}
\caption{The average absolute error (AAE) and average absolute error percentage (AAEP) of the heart rate estimation for the ICASSP 2015 signal processing cup database, where the data is downsampled to 25 Hz instead of the original 125 Hz. The datasets have been also truncated as proposed in \cite{Zhang2015}. The unit for the heart rate is beats per minute. Sbj: subject. Std: standard deviation.}
\label{tab:Motion_HR_AAEP_Truncated}
\end{minipage}
\end{table*}

\end{document}